\documentclass[12pt]{article}
\usepackage{feynarts,epsfig,amssymb,amsmath,psfrag}
\usepackage{cite}

\def\soverm{ \frac{s}{m^2} }
\def\toverm{ \frac{t}{m^2} }

\def\Ell{{(L)}}
\def\One{{(1)}}
\def\Two{{(2)}}
\def\Three{{(3)}}
\def\pel{{(\ell)}}
\def\cG{  {\cal G}  }
\def\tc {  \tilde{c}  }
\def\tf {  \tilde{f}  }
\def\tC {  \tilde{C}  }
\def\al{\alpha}
\def\bet{\beta}
\def\ga{\gamma}
\def\Gam{\Gamma}
\def\de{\delta}
\def\Ibox{ I_{\rm single~box} }
\def\Idbox{ I_{\rm double~box} }
\newcommand{\cN}{{\cal N}}
\newcommand{\cR}{{\cal R}}
\newcommand{\cO}{{\cal O}}

\def\eqn#1{eq.~(\ref{#1})} 
\def\Eqn#1{Equation~(\ref{#1})}
\def\eqns#1#2{eqs.~(\ref{#1}) and~(\ref{#2})}

\def \be  {\begin{equation}}
\def \ee  {\end{equation}}
\def \ba  {\begin{eqnarray}}
\def \ea  {\end{eqnarray}}

\makeatletter
\def\mr@ignsp#1 {\ifx\:#1\@empty\else #1\expandafter\mr@ignsp\fi}%
\newcommand{\multiref}[1]{\begingroup
\xdef\mr@no@sparg{\expandafter\mr@ignsp#1 \: }%
\def\mr@comma{}%
\@for\mr@refs:=\mr@no@sparg\do{\mr@comma\def\mr@comma{,}\ref{\mr@refs}}%
\endgroup}
\makeatother

\ifx\href\asklfhas\newcommand{\href}[2]{#2}\fi

\begin{document}

\begin{flushright}
HU-EP-10/17\\
BOW-PH-147\\
BRX-TH-618\\
Brown-HET-1594
\end{flushright}
\vspace{3mm}

\begin{center}
{\Large\bf\sf  
More loops and legs in Higgs-regulated $\mathcal{N}=4$ SYM amplitudes
\\
}

\vskip 5mm 
Johannes M. Henn$^{a}$, 
Stephen G. Naculich\footnote{Research supported in part by the 
NSF under grant PHY-0756518}$^{,b}$, 
Howard J. Schnitzer\footnote{Research supported in part 
by the DOE under grant DE--FG02--92ER40706}$^{,c}$
and Marcus Spradlin\footnote{Research supported in part
by the DOE under grant DE--FG02-91ER40688\\
{\tt 
henn@physik.hu-berlin.de, naculich@bowdoin.edu, schnitzr@brandeis.edu,
marcus\_spradlin@brown.edu}
}$^{,d}$
\end{center}

\begin{center}
$^{a}${\em Institut f\"ur Physik\\
Humboldt-Universit\"at zu Berlin, 
Newtonstra\ss{}e15, D-12489 Berlin, Germany}

\vspace{5mm}

$^{b}${\em Department of Physics\\
Bowdoin College, Brunswick, ME 04011, USA}

\vspace{5mm}

$^{c}${\em Theoretical Physics Group\\
Martin Fisher School of Physics\\
Brandeis University, Waltham, MA 02454, USA}

\vspace{5mm}

$^{d}${\em 
Department of Physics\\
Brown University, Providence, RI 02912, USA }

\end{center}

\vskip 2mm

\begin{abstract}

We extend the analysis of Higgs-regulated 
planar amplitudes of $\cN=4$ supersymmetric Yang-Mills theory 
to four loops for the four-gluon amplitude
and to two loops for the five-gluon amplitude.
Our calculations are consistent with a proposed 
all-loop ansatz for planar MHV $n$-gluon amplitudes
that is the analog of the BDS ansatz in dimensional regularization.
In all cases considered, 
we have verified that the IR-finite parts of the logarithm 
of the amplitudes have the same dependence on kinematic variables
as the corresponding functions in dimensionally-regulated amplitudes
(up to overall additive constants, which we determine).

We also study various Regge limits of $\cN=4$ SYM
planar $n$-gluon amplitudes.
Euclidean Regge limits of Higgs-regulated $n \geq 4$ amplitudes yield results 
similar in form to those found using dimensional regularization, 
but with different expressions for the gluon trajectory 
and Regge vertices resulting from the different regulator scheme.
We also show that the Regge limit of the 
four-gluon amplitude is dominated 
at next-to-leading-log order by vertical ladder diagrams together 
with the class of vertical ladder diagrams with a single H-shaped insertion.

\end{abstract}

\vfil\break

\section{Introduction}
\setcounter{equation}{0}

The assumption of dual conformal symmetry 
has proven useful for understanding the structure of 
the large-$N$ limit of higher-loop scattering amplitudes
in $\cN=4$ supersymmetric Yang-Mills theory.
Dual conformal invariance characterizes \cite{Drummond:2006rz} 
the set of scalar diagrams that 
forms a basis for the computation of higher-loop amplitudes 
in the unitarity-based approach of refs.~\cite{Bern:1994zx,Bern:1994cg,Buchbinder:2005wp,Bern:2007ct,Cachazo:2008dx}.
Moreover, coupled with the assumption of Wilson loop/MHV scattering amplitude 
duality
\cite{Alday:2007hr,Drummond:2007aua,Brandhuber:2007yx,Drummond:2007cf}, 
dual conformal symmetry has been used \cite{Drummond:2007au} to explain 
why maximally-helicity-violating planar $n$-gluon amplitudes
obey the Bern-Dixon-Smirnov (BDS) ansatz \cite{Bern:2005iz}
for $n=4$ and 5.
It also helps to explain why the BDS ansatz fails
to predict the scattering amplitude starting at $n=6$,
as observed in 
refs.~\cite{Alday:2007he,Bartels:2008ce,Bern:2008ap,Drummond:2008aq},
and constrains the form of the discrepancy, called the remainder function 
(for reviews see refs.~\cite{Alday:2008yw,Henn:2009bd}).
Wilson loop/MHV amplitude duality has 
been used to determine
the precise form of this 
remainder function for $n=6$ in
refs.~\cite{Drummond:2008aq,DelDuca:2009au,DelDuca:2010zg,Zhang:2010tr}.

Scattering amplitudes in massless gauge theories possess infrared
(IR)
divergences that must be regulated.
It is desirable that the regulator preserve as many symmetries as
possible. 
While dimensional regularization explicitly breaks dual conformal symmetry
(see,  e.g., ref.~\cite{Beisert:2010gn}), 
an alternative Higgs regulator 
proposed by Alday, Henn, Plefka, and Schuster \cite{Alday:2009zm} 
leaves the dual conformal symmetry unbroken.
In this approach, the supersymmetric
Yang-Mills (SYM) theory is considered on the Coulomb
branch where scalar vevs break the gauge symmetry, causing some 
of the gauge bosons to become massive through the Higgs mechanism.
Planar gluon scattering amplitudes on this branch can be computed using
scalar diagrams in which some of the internal and external states are
massive, regulating the IR divergences of the scattering amplitudes.
(For earlier applications of a massive IR regulator, see
refs.~\cite{Alday:2007hr,Korchemsky:1988pn,Kawai:2007eg,Schabinger:2008ah,McGreevy:2008zy};   
low-energy amplitudes in the Higgsed phase of SYM have been studied
in ref.~\cite{Gorsky:2009dr}.)
The diagrams remain dual conformal invariant, however, provided that
the dual conformal generators are taken to act on the masses 
as well as on the kinematical variables,    
a generalization
referred to as extended dual conformal symmetry.

The assumption of extended dual conformal symmetry\footnote{or 
rather the slightly stronger assumption 
(which holds in every known example) that every amplitude
can be expressed in an integral basis in which each element 
is dual conformal invariant  
}
severely restricts the number of diagrams that can appear
in the scattering amplitudes.
In particular it forbids loop integrals containing triangles,
which have indeed recently been shown to be absent from
one-loop amplitudes on the Coulomb branch~\cite{Boels:2010mj}
(a related discussion was given in ref.~\cite{Schabinger:2008ah}).
At one point in moduli space, 
all the lines along the periphery of the diagrams have mass $m$, 
while the external states and the lines in the interior of the diagram 
are all massless.  
This is believed to be sufficient to regulate all IR divergences of
planar scattering amplitudes.\footnote{More precisely,
diagrams which cannot be rendered finite in this manner
also cannot be rendered finite in off-shell regularization,
and it is believed that such diagrams 
never contribute to the amplitude
\cite{Drummond:2007aua}.
}
The original SYM theory is then recovered by taking $m$ small.

Using the Higgs regulator described above, 
the $\cN=4$ SYM planar four-gluon amplitude 
was computed at one and two loops in ref.~\cite{Alday:2009zm}
and at three loops in ref.~\cite{Henn:2010bk},
assuming that only integrals invariant
under extended dual conformal symmetry contribute.
It was shown that, at least through three loops,
the Higgs-regulated four-gluon amplitude
obeys an exponential ansatz
\ba
\label{intro-bds} 
\log M_4 (s,t) 
&=& 
-\, {1\over 8} \gamma(a) 
\left[ \log^2 \Bigl(\soverm \Bigr) + \log^2 \Bigl(\toverm\Bigr) \right]
- \tilde{\cG}_{0}(a) 
\left[ \log \Bigl(\soverm\Bigr) + \log \Bigl(\toverm\Bigr) \right]
\nonumber\\
&&+\, {1\over 8} \gamma(a) 
\left[ \log^2 \left(s\over t\right) + \pi^2\right] +{\tc_4}(a) + \cO(m^2) 
\ea
analogous to the BDS ansatz in dimensional regularization. 
Here $M_4 (s,t)$ is the ratio of the all-orders planar amplitude
to the tree-level amplitude,
$\gamma(a)$ is the cusp anomalous dimension \cite{Korchemskaya:1992je},
and $\tilde{\cG_0} (a)$ and $\tc_4 (a)$ 
are analogs of functions appearing in the BDS ansatz.
One of the advantages of the Higgs-regulated ansatz (\ref{intro-bds})
is that IR divergences take the form of logarithms of $m^2$;
consequently, the $L$-loop amplitude may be computed 
by simply exponentiating $\log M_4 (s,t)$ without regard 
for the $\cO(m^2)$ terms since they continue to vanish as 
$m \to 0$ even when multiplied by logarithms of $m^2$.  
Putting it another way, in order to test \eqn{intro-bds}
one need not compute any $\cO(m^2)$ terms of the 
Higgs-regulated $L$-loop amplitudes
because they cannot make any contribution to the 
IR-finite part of $\log M(s,t)$. 
This stands in stark contrast to dimensional regularization,
where checking the BDS ansatz at each additional loop order requires recalculation of two
more terms in the $\epsilon$ expansion of each lower loop integral.

One of the results of the current paper is to
extend the computation of the Higgs-regulated 
four-gluon amplitude to four loops.  
Following the observed behavior
through three loops, we assume that
the Higgs-regulated four-loop amplitude can be expressed as the same linear combination
of eight scalar integrals as the dimensionally-regulated amplitude.
Specializing to the kinematic point $s=t$, 
we demonstrate that the result is consistent 
with the exponentiation of IR divergences built into the BDS ansatz
(\ref{intro-bds}),
and we confirm, with significantly
improved numerical precision,
the value of the four-loop cusp anomalous dimension 
found in refs.~\cite{Bern:2006ew,Cachazo:2006az}.

Taking the Regge limit $s \gg t$,
the ansatz (\ref{intro-bds}) implies that the 
Higgs-regulated four-gluon amplitude exhibits exact Regge 
behavior \cite{Drummond:2007aua,Fadin:1996tb,Korchemskaya:1996je,Kotikov:2000pm,Naculich:2007ub,DelDuca:2008pj,Naculich:2009cv} 
\be
\label{intro-regge}
M_4 (s,t) = M_4 (t,s) = \beta (t) \left( \soverm\right)^{\alpha(t)-1}
\ee
where the all-loop-orders Regge trajectory is
\be
\label{intro-alpha}
\alpha(t) -1
= - {1 \over 4} \gamma(a) \log \Bigl(\toverm\Bigr) - \tilde{\cG}_{0} (a).
\ee
The coefficient of $\log t$, the cusp anomalous dimension, 
is independent of the IR regulator, 
while the constant part is scheme-dependent.
In ref.~\cite{Henn:2010bk}, 
we verified that the Higgs-regulated four-gluon amplitude 
obeys \eqn{intro-regge} through three-loop order,
and in the current paper, we extend this to four loops.

\Eqn{intro-bds} implies that the 
$L$-loop amplitude has leading log (LL) expansion
\ba
\label{intro-ll-expansion}
M_4^\Ell  
&=&
\frac{(-1)^L}{L!} 
\log^L  \Bigl(\toverm \Bigr)
\log^L  \Bigl(\soverm\Bigr) 
\nonumber\\
&+&
(-1)^{L-1}
\left[
	\left(
		\frac{\pi^2}{2(L-1)!} -\frac{\pi^2 }{6(L-2)!}
	\right) 
\log^{L-1}  \Bigl(\toverm \Bigr)
-                 \frac{\zeta_3}{(L-2)!}
\log^{L-2}  \Bigl(\toverm \Bigr)
\right]  
\log^{L-1}  \Bigl(\soverm\Bigr) 
\nonumber\\
&+&\cO\left( 
\log^{L-2}  \Bigl(\soverm\Bigr) 
\right)
\ea
where the LL $\log^L s$
term depends only on the lowest-order term
$\gamma(a) = - 4a + \cO(a^2)$ 
of the cusp anomalous dimension,
while the next-to-leading log (NLL)
$\log^{L-1} s$ term depends on the $\cO(a^2)$ terms
of $\gamma(a)$ and $\tilde{\cG}_0 (a)$.
We showed in ref.~\cite{Henn:2010bk} that the LL term
stems entirely from a single scalar diagram, 
the vertical ladder, 
in the Regge limit of the Higgs-regulated loop expansion.\footnote{
\label{fn}
There are two ways of taking the Regge limit of a Higgs-regulated
amplitude.
One can either 
(a) first take the limit $m^2 \ll s,t$,
and then $s \gg t$, or 
(b) first take the  limit $s \gg t$, $m^2$, and then $m^2 \ll t$. 
We demonstrated in ref.~\cite{Henn:2010bk} that
the amplitude is independent of the order of limits,
at least through three loops.
The Regge behavior of individual diagrams, however, can depend on 
the order in which the limits are taken.   
The dominance of vertical ladder diagrams is only valid in the
Regge (b) limit, 
and the discussion in the text assumes this order of limits.
}
In this paper, we show that the NLL contribution 
to the $L$-loop amplitude
is given by the subleading term of the vertical ladder, 
together with the leading contributions from a set of $(L-1)$ diagrams,
consisting of vertical ladder diagrams with a single 
H-shaped insertion.
We explicitly confirm this through five loops
via a Mellin-Barnes computation,
and present an argument (subject to certain reasonable assumptions)
for its validity to all loops.

A second thrust of this paper is to consider 
higher-point amplitudes in Higgs regularization.
We compute the Higgs-regulated five-gluon amplitude 
at one- and two-loops and 
establish 
the iterative relation 
\be
\label{intro-two-loop}
M_5^{(2)} = \frac{1}{2} \left( M_5^{(1)} \right)^2 
+
\sum_{i=1}^5 
\left[
\frac{\zeta_2}{4} 
\log^2\left(\frac{s_i}{m^2} \right) 
+ \frac{\zeta_3}{2} \log \left( \frac{s_i}{m^2} \right)
\right]
 - \zeta_2  \, F_5^\One (s_i) + 
{5 \over 4} \zeta_4 
+ \cO(m^2)
\ee
where
\be
F_5^\One (s_i)  = \lim_{m^2 \to 0} \left[
M_5^\One + \frac{1}{4} \sum_{i=1}^5
\log^2 \left( \frac{s_i}{m^2} \right)
\right]
\ee
is the IR-finite part of the one-loop amplitude, which is the same
as in dimensional regularization, up to an additive constant.
We argue that the parity-odd part of $M_5^\Two$ is 
at most $O(m^2)$ in Higgs regularization.

Based on the iterative relation (\ref{intro-two-loop}), 
we propose that the 
generalization of \eqn{intro-bds} to the planar MHV $n$-gluon amplitude 
in Higgs regularization takes the form 
\be
\label{intro-BDSnparticle}
\log M_n = \sum_{i=1}^n
\left[ - \frac{\gamma(a)}{16} 
\log^2 \left( \frac{s_i}{m^2} \right)
- \frac{\tilde{\cal G}_0(a)} {2}
\log\left( \frac{s_i}{m^2} \right)
+ \tf (a)
\right]
+ \frac{1}{4} \gamma(a) \, F_n^{(1)} + \cR_n + \tC (a)
+ \cO(m^2)
\ee
with 
\be
\tf(a)  =  \frac{\pi^4}{180} a^2 + {\cal O}(a^3),
\qquad\qquad 
\tC(a) = - \frac{\pi^4}{72} a^2 + {\cal O}(a^3)
\ee
and $\cR_n$ vanishes for $n=4$ and $n=5$.
For $n\ge 6$, we expect the
remainder function $\cR_n$
to be equal to its counterpart in dimensional regularization.

Various Regge limits of the $n$-gluon amplitude for $n \geq 5$
can be defined (see, e.g., refs.~\cite{Brower:2008nm,  Brower:2008ia}). 
We consider the single and double Regge limits (in the Euclidean region)
of the Higgs-regulated five-gluon amplitude up to two loops.
In both of these cases, the double logarithm in $\log M_5$
cancels out, leaving single logarithmic dependence on 
the large kinematic variable.

In ref.~\cite{Henn:2010bk},  
an alternative approach to the Regge limit for four-gluon amplitudes
was considered by taking a different point on the Coulomb branch,
involving two different masses.
We extend this approach to single and double Regge limits of 
the five-gluon amplitude.
In both cases, this alternative approach 
makes clear that $\log M_5$ should only have single logarithmic 
dependence on  the kinematic variables
because collinear divergences are absent.
A similar approach can be taken for some, 
but not necessarily all, of the  $n=6$ 
Regge limits considered in refs.~\cite{Brower:2008nm,Brower:2008ia}.

The paper is organized as follows.
In section 2 we study the four-gluon amplitude, presenting
explicit results for all contributing four-loop
integrals at the symmetric point
$s = t$ and arguing that, to all loops,
the NLL contribution to the amplitude in the
Regge limit is given by a small subset of all diagrams.
In section 3 we turn our attention to the five-gluon amplitude
at two loops, the evaluation of which leads
to \eqn{intro-two-loop}.
Section 4 contains a discussion of various Regge limits of Higgs-regulated
amplitudes for $n \ge 5$, with emphasis on those
features which differ from similar limits
of dimensionally-regulated amplitudes.
Section 5 summarizes our results, while various technical 
details can be found in three appendices.

\section{The four-point amplitude}
\setcounter{equation}{0}

In ref.~\cite{Alday:2009zm}, it was suggested that the analog
of the Bern-Dixon-Smirnov ansatz \cite{Bern:2005iz} 
for the planar four-point amplitude in Higgs regularization is
\ba
\label{all-loop-higgs}
\log M_4(s,t) 
&=&
-\, \frac{1}{8} \gamma(a)
\left[ \log^2 \Bigl(\soverm \Bigr) + \log^2 \Bigl(\toverm\Bigr) \right]
- \tilde{\cG}_{0}(a)
\left[ \log \Bigl(\soverm\Bigr) + \log \Bigl(\toverm\Bigr) \right]
\nonumber\\
&&+\, \frac{1}{8} \gamma(a)
\left[ \log^2 \left(\frac{s}{t}\right) + \pi^2\right] +{\tc_4}(a) + \cO(m^2)
\ea
where
\be
\label{defgamma}
\gamma(a)
= \sum_{\ell=1}^\infty a^\ell \gamma^\pel
=
4a - 4\zeta_2 a^2 + 22 \zeta_4 a^3 + \cO(a^4)
\ee
is the cusp anomalous dimension, and
\be
\label{twoloopanom}
\tilde{\cG}_{0} (a) = - \zeta_{3} a^2  + \cO(a^3), \qquad
\tc_{4}(a) = \frac{\pi^4}{120} a^2 + \cO(a^3) 
\ee
are analogs of functions appearing in the BDS ansatz in 
dimensional regularization \cite{Bern:2005iz}, 
but need not be identical since they are 
scheme-dependent \cite{Alday:2009zm}.
Overlapping soft and collinear IR
divergences
are responsible for the double logarithms in \eqn{all-loop-higgs}.
The nontrivial content of \eqn{all-loop-higgs} is the
statement about the finite terms; 
the IR singular terms of the amplitude
are expected to obey \eqn{all-loop-higgs} on general field theory grounds
(see refs.~\cite{Catani:1998bh,Sterman:2002qn}).

According to the assumption of dual conformal symmetry, 
the planar $L$-loop amplitude can be written as
\begin{equation}\label{dcs-conjecture}
M^{(L)} =  \sum_{ {I}  } \, c({I} ) \,{I} 
\end{equation}
where the sum runs over all extended dual conformal integrals ${I}$,
with some coefficients $c({I} )$.
In the case of the four-point amplitude, 
the coefficients are simply numbers.
The set of loop integrals invariant under extended dual conformal symmetry
is significantly smaller than that of generic loop integrals.

At two and three loops,
the assumption (\ref{dcs-conjecture}),
together with the infrared consistency conditions, 
leads to a result in agreement with the exponential ansatz 
(\ref{all-loop-higgs}),
with the values (\ref{defgamma}), (\ref{twoloopanom}) 
and\footnote{In the course of
the four-loop computation of the present paper 
we have improved the numerical accuracy of 
$\tilde{\cG}_{0}^{(3)}$ and $\tc_4^{(3)}$ quoted in ref.~\cite{Henn:2010bk}.
Here we display the results with improved numerical accuracy.}
\cite{Henn:2010bk}
\be
\label{threeloopanom}
\tilde{\cG}_{0}^{(3)} \approx 2.688870547851
\pm 6.5 \times 10^{-11}, \qquad
\tc_4^{(3)} \approx -9.24826993 \pm 9.6 \times 10^{-7}\,.
\ee
If one assumes that these coefficients have transcendentality\footnote{If 
we attribute
a degree of transcendentality $0,1,1$ and $n$, respectively
to rational numbers, $\pi$, $\log$ and $\zeta_{n}$,
and define the transcendentality of a product to be additive, 
then the $L$-loop amplitude is expected to have
uniform transcendentality $2L$.}
5 and 6 respectively,
one finds for them the probable analytic values\footnote{We thank
Lance Dixon for suggesting this value of $\tilde{\cG}_0^{(3)}$ to us.}
\be
\label{pslq}
\tilde{\cG}_{0}^{(3)} = \frac{9}{2} \zeta_5 - \zeta_2  \zeta_3  ,
\qquad
\tc_4^{(3)} = - \frac{25}{4} \zeta_6 - 2 \zeta_3^2
\ee
using the PSLQ algorithm \cite{Ferguson91apolynomial,PSLQ}.

The three-loop amplitude was computed \cite{Henn:2010bk} 
by assuming that 
\be
M_4^\Three(s,t)  
= 
- \frac{1}{8} \, \Big[  I_{3a}(s,t) + 2 \, I_{3b}(s,t) \Big]
+ (s \leftrightarrow t) \,.
\ee
There are two additional dual conformal invariant three-loop integrals,
$I_{3c}$ and $I_{3d}$,
with powers of $m^2$ in the numerator, 
but which have a finite $m\to 0$ limit.
These could in principle contribute to the three-loop amplitude
if the coefficients multiplying them are nonzero.
In ref.~\cite{Henn:2010bk}, we showed 
that even in this case
the exponential ansatz remains valid, 
provided that the coefficients 
$\tilde{\cG}_{0}^{(3)} $ and $ \tc_4^{(3)}$ 
are shifted accordingly.

\begin{figure}[t]
 \centerline{
 {\epsfxsize15cm  \epsfbox{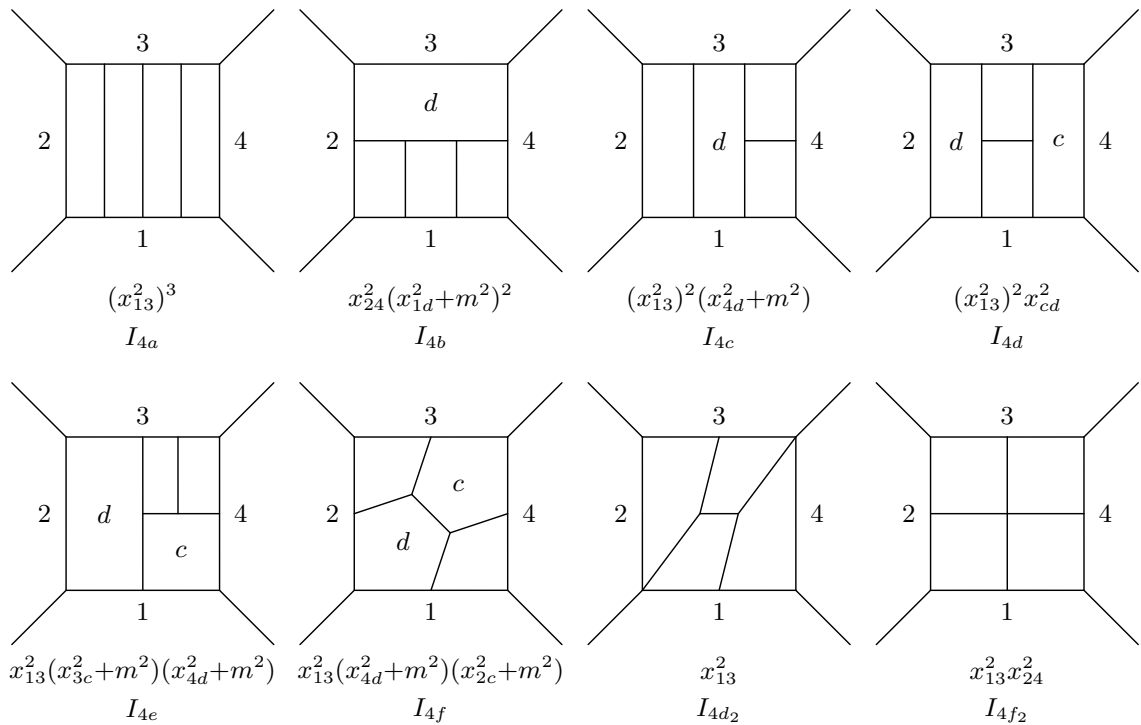}} 
}
\caption{\small 
The eight diagrams contributing to the four-loop 
four-point amplitude.
We use the standard dual variable notation, 
labeling the external faces by $x_1$ through $x_4$ 
and the internal faces by $x_a$ through $x_d$.  
The former are related to the external momenta via 
$p_i = x_i  - x_{i+1} $ (where $i$ is understood mod 4) while the
latter are each integrated with the measure $d^4 x/(i \pi^2)$.
Under each diagram is shown the numerator factor for
the corresponding integral.
To avoid clutter, 
we omit an overall factor of $x_{13}^2 x_{24}^2 = s t$ from
each diagram (where $x_{ab} \equiv x_a - x_b$),
and we do not label internal faces 
not appearing in numerator factors.
As an illustrative example we demonstrate how to assemble all ingredients
of the integral $I_{4b}$ in \eqn{twopointnine}.}
\label{fig-four-loop}
\end{figure}

\subsection{Four-loop four-point amplitude}

For the four-loop four-gluon amplitude, we will use the 
ansatz \cite{Bern:2006ew}
\begin{eqnarray}\label{M4}
M_{4}^{(4)}(s,t) &=&\frac{1}{16}
\Big[ I_{4a}(s,t) + 2 I_{4b}(s,t) + 2 I_{4c}(s,t)
+ I_{4d}(s,t) + 4 I_{4e}(s,t) + 2 I_{4f}(s,t)  \nonumber \\
&& \phantom{spa}- 2 I_{4 d_{2}}(s,t) - \frac{1}{2} I_{f_{2}}(s,t)
\Big] + (s \leftrightarrow t)
\end{eqnarray}
where the individual integrals,
shown in fig.~\ref{fig-four-loop},
are defined in dimensional regularization in ref.~\cite{Bern:2006ew}.
These integrals are all dual conformal invariant, 
and are straightforwardly rewritten in Higgs regularization, following 
refs.~\cite{Alday:2009zm,Henn:2010bk}.
For example, using dual coordinates \cite{Drummond:2006rz}
for convenience, we have
\begin{align}
\label{twopointnine}
I_{4b}(s,t)  &= \int \frac{d^{4}x_{a} d^{4}x_{b} d^{4}x_{c} d^{4}x_{d}}{ (i \pi^2)^4} \frac{x_{13}^2 (x_{24}^2)^2 (x_{1d}^2 + m^2 )^2}{(x_{1a}^2 +m^2 ) (x_{1b}^2 +m^2 ) (x_{1c}^2 +m^2 ) 
(x_{2a}^2 +m^2 ) (x_{2d}^2 +m^2 ) }  \nonumber \\
& \qquad \times \frac{1}{(x_{3d}^2 +m^2 ) (x_{4d}^2 +m^2 ) (x_{4c}^2 +m^2 ) x_{ab}^2 x_{bc}^2 x_{ad}^2 x_{bd}^2 x_{cd}^2 } \,.
\end{align}
Note that in $I_{4d}$ there is no $+m^2$ term in the loop-dependent numerator, 
since it connects two internal integration points.

We have written down Mellin-Barnes representations for all 
eight integrals. This is easily done introducing the MB
representation loop by loop \cite{Smirnov:2006ry, Henn:2010bk}.
Interestingly, the dimensionality of most MB representations
is one lower than the corresponding representation 
in dimensional regularization. At
two and three loops, the opposite was the case.

To make contact with the exponential ansatz (\ref{all-loop-higgs}),
we will compute $M_{4}(s,t)$ at the symmetric point $s=t$.
Defining $x = m^2/s = m^2/t$, the expression above becomes
\be\label{M4x}
M_{4}^{(4)}(x) =\frac{1}{16}
\Big[ 2 I_{4a}(x)+ 4 I_{4b}(x) + 4 I_{4c}(x) + 2 I_{4d}(x) +8 I_{4e}(x) +4 I_{4f}(x) - 4 I_{4 d_{2}}(x)  - I_{f_{2}}(x)
\Big]\,.
\ee
To evaluate these integrals, we proceed as in 
refs.~\cite{Alday:2009zm,Henn:2010bk}.
The starting point is a multi-dimensional MB representation depending
on the parameter $x$. For simplicity, consider a one-dimensional MB integral
\begin{align}
\int_{\beta -i \infty}^{\beta + i \infty} \, dz \,x^{z}\, f(z)
\end{align}
where typically $f(z)$ is a product of $\Gamma$ functions, and $\beta<0$. 
In principle, one could close the 
integration contour, say, on the right and obtain the answer as an infinite series 
arising from poles of the $\Gamma$ functions in $f(z)$.
However, since we are only interested in the $\log x$ terms as $x \to 0$,
it is sufficient to deform the integration contour to positive values of ${\rm Re}(z)$.
The logarithms arise from taking residues at $z=0$.
In the case of multi-fold MB integrals, the above strategy can be iterated.
We obtain expressions of the form 
$\sum_{i=0}^{6} b_{i} \, \log^{8-i} x + \cO( \log x )$
using the Mathematica code {\it MBasymptotics} \cite{MBasymptotics}.
In general, the coefficients $b_{i}$ 
still involve a (significantly lower) number
of MB integrals, which we evaluate numerically using 
the code {\it MB} \cite{Czakon:2005rk}.
Denoting $L = \log x$, we find
\begin{eqnarray}
I_{4a} (x) &=&  \frac{1}{56}  L^8 +  \frac{8}{135} \pi^2 L^6 -\frac{8}{15} \zeta_{3} L^5  -\frac{2}{27} \pi^4 L^4   \\
             && + \left( -\frac{32}{3} \zeta_{2} \zeta_{3} - \frac{8}{3} \zeta_{5} \right) L^3 
              +( -162.26621838645 \pm  1.9 \times 10^{-10}  ) L^2 + \cO(L) \,. \nonumber \\
   I_{4b} (x) &=& \frac{149}{5040} L^8 - \frac{1}{60} \pi^2 L^6 -\frac{6}{5} \zeta_{3} L^5 +  \frac{11}{72} \pi^4 L^4 
     \\   &&  + \left( \frac{151}{3} \zeta_{2} \zeta_{3} + \frac{247}{6} \zeta_{5} \right) L^3 
 + (525.46852427784 \pm 9.9 \times 10^{-10} ) L^2 + \cO(L) \,.  \nonumber \\
             I_{4c} (x) &=& \frac{271}{10080} L^8  -\frac{2}{3} \zeta_{3} L^5 -  \frac{11}{270} \pi^4 L^4 \\
&&             + \left(-2 \zeta_{2} \zeta_{3} - \frac{127}{3} \zeta_{5}  \right) L^3 
 +( -128.86933736 \pm 4.2 \times 10^{-7} ) L^2 + \cO(L) \,.\nonumber\\
 I_{4d} (x) &=& \frac{9}{560} L^8  + \frac{19}{270} \pi^2 L^6  -\frac{4}{15} \zeta_{3} L^5 -  \frac{19}{135} \pi^4 L^4 \\
 &&+ \left(-\frac{128}{3} \zeta_{2} \zeta_{3} - 40  \zeta_{5}  \right) L^3
  +( -710.51212126801 \pm 1.4 \times 10^{-10} ) L^2 + \cO(L) \,. \nonumber \\
 I_{4e} (x) &=& \frac{271}{10080} L^8  + \frac{1}{60} \pi^2 L^6  -\frac{2}{3} \zeta_{3} L^5 -  \frac{191}{2160} \pi^4 L^4\\
 &&
  + \left(-\frac{23}{6} \zeta_{2} \zeta_{3} - \frac{335}{12}  \zeta_{5}  \right) L^3
   + (222.7007725 \pm 1.8 \times 10^{-6} ) L^2 + \cO(L) \nonumber \,.
\end{eqnarray}
\begin{eqnarray}
  I_{4f} (x) &=& \frac{199}{2520} L^8  - \frac{22}{135} \pi^2 L^6  -\frac{12}{5} \zeta_{3} L^5 +  \frac{17}{30} \pi^4 L^4 \\
  && + \left( 68 \zeta_{2} \zeta_{3} + \frac{286}{3}  \zeta_{5}  \right) L^3
 + ( -117.32774717 \pm 1.8 \times 10^{-7} ) L^2 + \cO(L) \,. \nonumber\\
   I_{4 d_{2}} (x) &=&     -\frac{8}{15} \zeta_{3} L^5 +  \frac{2}{45} \pi^4 L^4 \\ 
   && + \left( 16 \zeta_{2} \zeta_{3} - \frac{16}{3}  \zeta_{5}  \right) L^3
 + (180.37203096920 \pm 6.6 \times 10^{-10})  L^2 + \cO(L) \,. \nonumber\\
  I_{4 f_{2}} (x) &=& \frac{199}{1260} L^8  - \frac{44}{135} \pi^2 L^6  -\frac{88}{15} \zeta_{3} L^5 +  \frac{17}{15} \pi^4 L^4 \\
  && + \left( 168 \zeta_{2} \zeta_{3} + \frac{700}{3}  \zeta_{5}  \right) L^3
 +( 324.1906414642  \pm 2.6 \times 10^{-9} )  L^2 + \cO(L) \nonumber \,.
\end{eqnarray}
Summing up the contributions of the four-loop integrals computed above
using \eqn{M4x},
we obtain
\begin{equation}
\label{M4result}
M_{4}^{(4)}(x) = \frac{1}{24} L^8 - \zeta_{3} L^5 + \frac{1}{60} \pi^4 L^4 + \left( 6 \zeta_{2} \zeta_{3} - 9 \zeta_{5} \right) L^3 + (6.71603090 \pm 9.1 \times 10^{-7}) L^2  + \cO(L)\,.
\end{equation}
The coefficients of $L^4$ and $L^3$ in the expressions above 
were obtained numerically,
so we cannot distinguish between rational and transcendental numbers. 
Nevertheless,
motivated by the expectation that the result should have uniform 
transcendentality,
we have replaced the numerical values by their  probable analytical equivalents.
Specifically,
we used the Mathematica implementation of the PSLQ algorithm 
to identify linear combinations of numbers with 
the correct degree of transcendentality 
that agree with our results within the numerical accuracy. 
It goes without saying that this does not constitute a proof that 
these expressions are necessarily correct.
A guess for the numerical coefficient of the $L^2$ term is
$-\frac{1}{2}\zeta_{6} + 5 \zeta_{3}^2$.

Exponentiation of the
IR logarithms requires that
\begin{eqnarray}\label{M4IRconsistency}
M_{4}^{(4)}(x) &=& \frac{1}{24} L^8 - \zeta_{3} L^5 + \frac{1}{60} \pi^4 L^4  + \left( 4 \zeta_{2} \zeta_{3} - 2 \tilde{\mathcal{G}}_{0}^{(3)} \right) L^3  \nonumber \\
&& + \left( - \tc_4^{(3)}  - \frac{1}{4} \gamma^{(4)} - \frac{13}{360} \pi^6 + 2 \zeta_{3}^{2} \right) L^2 + \cO(L) 
\end{eqnarray}
where we have used the values of
$\gamma(a)$, $\tilde{\cG}_{0} (a)$, and $\tc_4(a) $
given in \eqns{defgamma}{twoloopanom}. 
There is complete agreement between \eqns{M4result}{M4IRconsistency}.
Comparing the coefficients of $L^3$,
we confirm the value
$\tilde{\cG}_{0}^{(3)} = \frac{9}{2} \zeta_5 - \zeta_2  \zeta_3 $
obtained at three loops by 
assuming that $I_{3d}$ does not contribute.
Comparing the coefficients of $L^2$, 
we confirm the value 
$\tc_4^{(3)} = - \frac{25}{4} \zeta_6 - 2 \zeta_3^2$
obtained at three loops by assuming that $I_{3c}$ does not contribute,
provided that the four-loop 
cusp anomalous dimension is given by
\begin{align}
\label{f4}
\gamma^{(4)} = -117.1788222 \pm 3.7 \times 10^{-6}  \approx 
- 4 \zeta_{2}^3 -24  \zeta_{2} \zeta_{4} -4 \zeta_{3}^2 - 50 \zeta_{6} 
\end{align}
in perfect agreement with the result (also numerical) 
found in refs.~\cite{Bern:2006ew,Cachazo:2006az}, and
in agreement with the spin chain prediction from ref.~\cite{Beisert:2006ez}.
It is noteworthy that our result~(\ref{f4})
improves the numerical precision by two orders of magnitude
compared to ref.~\cite{Cachazo:2006az} and by five orders of magnitude
compared to ref.~\cite{Bern:2006ew}\footnote{The value~(\ref{f4}) corresponds
to $r = - 1.99999892 \pm 6.3 \times 10^{-7}$ in the parameterization used in
those references.}.

Similar to the situation at three loops,
there are $10$ four-loop integrals (in addition to those 
shown in \eqn{M4}) 
that could in principle contribute to the four-loop amplitude, 
based solely on requiring dual conformal invariance \cite{Nguyen:2007ya}.  
As we have seen, however, our results are
completely consistent with the absence of these additional 
integrals at three and four loops.

It is amusing that the coefficients appearing in the results for the
individual integrals are rather complicated (e.g.  the coefficients
of the $L^8$ terms), yet they sum up to give the very simple result
(\ref{M4result}), as required by infrared consistency. 
This suggests there may exist 
a better organization of the calculation that avoids
the complexity of the intermediate results.

\subsection{Regge limit of the four-point function}
\label{sect-regge-fourpoint}

The four-point amplitude (\ref{all-loop-higgs}) can be rewritten as
\be
\label{higgsregge}
\log M_4  
=
- \frac{1}{4} \gamma(a) (\log v)( \log u)
+ \tilde{\cG}_{0} (a)
\left( \log u + \log v \right)
+
\frac{\pi^2}{8} \gamma(a)
+ \tc_4 (a)  + \cO(m^2)
\ee
where $u = m^2/s$ and $v = m^2/t$.
Note that $(\log u)^2$ and $(\log v)^2$ terms are absent,
which is related to the Regge-exactness of the four-point amplitude.
This can alternatively be interpreted as follows \cite{Henn:2010bk}.
The Higgs-regulated amplitude can be computed at a 
different point in moduli space,
where the diagrams contain internal lines with 
different masses $m$ and $M$ along the periphery,
and external lines with mass $M-m$.     
By dual conformal invariance, the amplitude only depends on
$u = m^2/s$ and $v = M^2/t$.
The absence of $(\log m^2)^2$ terms in the amplitude
can be understood as the absence of collinear divergences 
when massive ($M$) particles scatter by exchanging lighter ($m$) 
particles
(with the mass of the lighter particles serving as an IR regulator).

\begin{figure}
 \psfrag{A}[cc][cc]{$(a)$}
\psfrag{B}[cc][cc]{$(b)$}
\psfrag{s}[cc][cc]{$s$}
\psfrag{t}[cc][cc]{$t$}
\psfrag{p1}[cc][cc]{$p_{1}$}
\psfrag{p2}[cc][cc]{$p_{2}$}
\psfrag{p3}[cc][cc]{$p_{3}$}
\psfrag{p4}[cc][cc]{$p_{4}$}
\psfrag{dots}[cc][cc]{$\dots$}
\psfrag{regge}[cc][cc]{$s \gg t$}
\psfrag{conj}[cc][cc]{}
\psfrag{ILa}[cc][cc]{$I_{L\,a}$}
\psfrag{ILH}[cc][cc]{$I_{L\,H}$}
\psfrag{LLandNLL}[cc][cc]{LL and NLL }
\psfrag{NLL}[cc][cc]{NLL}
 \centerline{
 {\epsfxsize15cm  \epsfbox{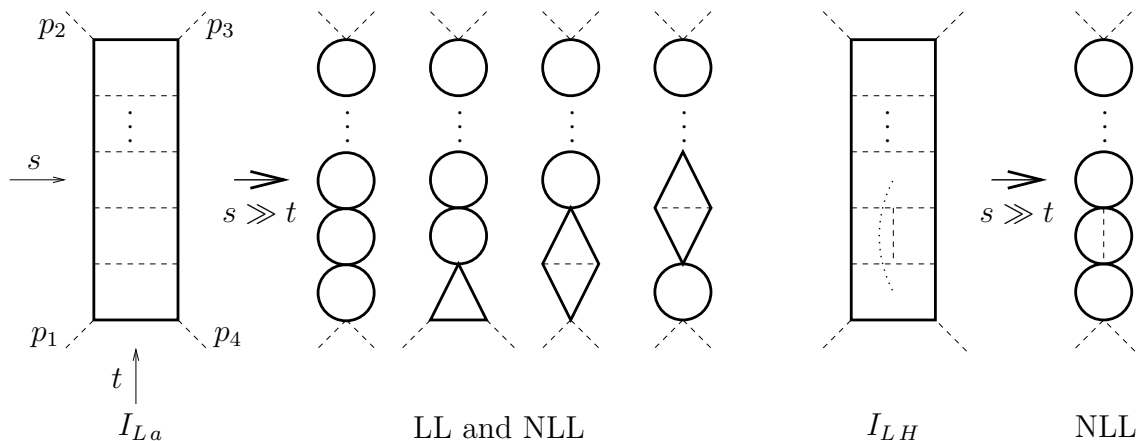}}
}
\caption{\small
Factorization of the leading-log and next-to-leading-log contributions 
to the Regge limit $s \gg t$ 
of the $L$-loop vertical ladder integral $I_{L\,a}(v,u)$
into simpler integrals. 
Factorization of the NLL contribution of the
vertical ladder integral with H-shaped insertion $I_{L\,H}$. 
The dotted line indicates a loop-momentum-dependent numerator.}
\label{fig-regge1}
\end{figure}

We may exponentiate \eqn{higgsregge},
and expand the result in powers of the coupling $a$ 
and of the Regge logarithms $\log u$.
At $L$-loop order, 
the leading logarithm is $\log^L u$, 
NLL is $\log^{L-1} u$, etc.
To LL and NLL order, the $L$-loop amplitude in the Regge limit 
is given by
\ba
\label{ll-expansion}
M_4^\Ell  
&=&
\left[
\frac{1}{L!} (-\log v)^L
\right]  \log^L u
\label{Lloop}
\nonumber\\
&+&
\left[
	\left(
		\frac{\pi^2}{2(L-1)!} -\frac{\pi^2 }{6(L-2)!}
	\right) (-\log v)^{L-1}
-                 \frac{\zeta_3}{(L-2)!}
         (-\log v)^{L-2}
\right]  \log^{L-1}  u
\nonumber\\
&+&\cO( \log^{L-2}  u )
\ea
where we have used \eqns{defgamma}{twoloopanom}. 
In contrast to dimensional regularization, 
in Higgs regularization there is a single diagram, 
the vertical ladder $I_{La}(v,u)$ (see fig.~\ref{fig-regge1}), 
that contributes\footnote{in the Regge (b) limit (see footnote \ref{fn}).}
to the LL term of the $L$-loop amplitude (\ref{ll-expansion}).
In the LL limit, the vertical ladder factorizes into a 
product of (two-dimensional) bubble integrals 
(again see fig.~\ref{fig-regge1}).
Moreover, the LL, NLL, and NNLL contributions of the vertical ladder diagram
were computed  (cf. eq.~(4.16) in ref.~\cite{Henn:2010bk})
using the method of ref.~\cite{Eden}. 
Subtracting the vertical ladder contribution from the prediction 
(\ref{ll-expansion}) for the full amplitude, we obtain
\be
\label{diff-regge-ladder}
 M_4^\Ell
- \left(-\frac{1}{2} \right)^L I_{La}  (v,u)
= \frac{(-1)^L }{3(L-2)!} \log^{L-1} u \left[ \log^{L+1} v + \pi^2 \log^{L-1} v + \cO(v) \right]
+ \cO(\log^{L-2} u )  \,.
\ee
{}From this, one sees that contributions from diagrams 
other than the vertical ladder
are required at NLL order to obtain the expected amplitude in
the Regge limit. 

Rules for evaluating the leading $\log s$ behavior of multiloop integrals 
were summarized in refs.~\cite{Eden,Collins:1977jy}.
One begins by identifying paths through the graph, 
which, when contracted to a point, 
split the diagram into two parts 
with a single vertex in common, 
and with $p_1$ and $p_4$ on one side and $p_2$ and $p_3$ on the other
(for example, each of the rungs of the vertical ladder diagram).
Paths of minimal length are called
``d-lines'' \cite{Halliday:1963xp} or ``t-paths'' \cite{Tiktopoulos:1963zz}.
A scalar diagram containing $m$ d-lines of length $n$ 
goes as $\log^{m-1}  s / s^n$ as $s \to \infty$. 
For example, the $L$-loop vertical ladder diagram
contains $(L+1)$ d-lines of length one, 
so the vertical ladder integral
(multiplied by $s t^L$ to make it dual conformal invariant)
goes as $\log^L s$, or equivalently $\log^L u$, as discussed above.
All other $L$-loop diagrams contain at most $(L-1)$ 
d-lines and hence {\it prima facie} 
give at most a $\log^{L-2} u$ contribution.
For example, 
the four-loop diagrams $I_{4c}(v,u)$ and $I_{4d}(v,u)$
(see fig.~\ref{fig-four-loop})
each contain two d-lines of length one
and {\it prima facie} go as $\log u$.
(These diagrams also contain two paths of length two, which
are not minimal and therefore do not contribute.)

The MB calculation summarized in appendix \ref{app-four-loop},
however, shows that both of these diagrams go as $\log^3 u$, 
two powers higher than expected.
This is because the d-line rules given above only apply 
to scalar diagrams with no non-trivial
(i.e., loop-momentum-dependent) numerator factors. 
The presence of numerator factors,
which are required for the loop integrations next to the 
H-shaped insertion to have the correct dual conformal weight,
can increase the leading power of $\log u$ of the diagram.
The reader may ascertain from the results of appendix \ref{app-four-loop}
that diagrams with no nontrivial numerator factors have 
the $\log u$ dependence predicted by the d-line rules,
whereas those with numerator factors can have a stronger $\log u$
dependence.

We show in appendix \ref{app-H} that
when an H-shaped insertion in a vertical ladder diagram
is accompanied by a numerator factor, 
the two lines of length two constituting the sides of the H
are effectively promoted to length one,
increasing the d-line count by two.
In particular, an $L$-loop vertical ladder diagram with a 
single H-shaped insertion $I_{LH}$
(see fig.~\ref{fig-regge1}),
which {\it prima facie} would go as $\log^{L-3} u$, 
actually goes as $\log^{L-1} u$
due to its numerator factors,
and thus contributes to the amplitude at NLL order.
A calculation in appendix \ref{app-H} further shows that,
subject to reasonable assumptions, 
the leading log contribution of the integral $I_{LH}$
factorizes as 
\begin{equation}
\label{LH}
 I_{L\,H} =   \frac{(-1)^{L-1} }{(L-1)!}  \log^{L-1} u \times
 K(v)^{L-2} \times K'(v)
   \quad  +\quad  \cO(\log^{L-2} u ) \,,
\end{equation}
where $K(v)$ and $K'(v)$ correspond to the two-dimensional bubble
and two-loop bubble diagrams shown in fig.~\ref{fig-regge1}  (see
ref.~\cite{Henn:2010bk} for further discussion).
Taking $v$ small, we have
\ba
K(v) &=& -2 \log v + O(v)\,, 
\nonumber\\
K'(v) &=& -\frac{4}{3} \log^{3} v -  \frac{4}{3} \pi^2 \log v + \cO(v) \,.
\ea
Note that \eqn{LH} implies that the position where
the H-shaped insertion is made into the vertical ladder integral is
unimportant at NLL order, i.e. all such integrals give the same NLL
contribution.

At two loops, $I_{LH}$ is just the horizontal ladder $I_{2a}(u,v)$.
At three loops, $I_{LH}$ is the tennis court diagram $I_{3b}(u,v)$.
In ref.~\cite{Henn:2010bk},
it was shown that the leading contribution of 
these diagrams is precisely given by \eqn{LH}.
At four loops, 
the three $I_{LH}$ diagrams are given by $I_{4c}(v,u)$, 
its flipped version, and $I_{4d}(v,u)$.
The results for these integrals
found in appendix \ref{app-four-loop}
are also in agreement with \eqn{LH} and,
moreover, all other four-loop  diagrams contribute at most to NNLL.
Finally, we have verified \eqn{LH} at five loops as well using the
integrals given in ref.~\cite{Bern:2007ct}.

At $L$-loop order there are $(L-1)$ vertical ladder diagrams 
containing a single $H$-shaped insertion, 
so that the total contribution to the amplitude of the vertical ladders
with one H-shaped insertion is
(multiplying the contributions by $(-1/2)^L$) 
\be
\left( - {1\over 2} \right)^L (L-1)  I_{L\,H} =   
\frac{(-1)^L }{3(L-2)!} \log^{L-1} u \left[ \log^{L+1} v + \pi^2 \log^{L-1} v + \cO(v) \right]
+ \cO(\log^{L-2} u )  
\ee
which precisely matches the result (\ref{diff-regge-ladder}) expected from the
exponential ansatz.

In summary, we have shown that in Higgs regularization, the
NLL contribution to the
four-gluon amplitude in the Regge limit 
is given by a small set of diagrams: the vertical ladders
and the vertical ladders with one H-shaped insertion.  We confirmed this
through five loops by direct evaluation of the integrals,
and we gave an argument that this holds to all loop orders.

\section{The five-point amplitude}
\label{sect-five-point}
\setcounter{equation}{0}

In this section, we direct our attention
to the two-loop $n=5$ point amplitude of $\cN=4$ SYM theory,
which is of interest 
for several reasons.  
It serves as a further instructive application 
of the Higgs mechanism to regulate infrared divergences, 
and confirms the universality of the 
exponential structure of IR singularities 
of $n$-point amplitudes in massless gauge theories
(which is usually studied in dimensional regularization;
for example, see refs.~\cite{Catani:1998bh,Sterman:2002qn}).
We also establish an iterative relation at two loops, 
which is the exact analog of the five-point iterative relation 
in dimensional regularization \cite{Cachazo:2006tj, Bern:2006vw}.
This allows us to write an all-loop ansatz
for $n$-point amplitudes in Higgs regularization,
analogous to the BDS ansatz in dimensional regularization,
whose content is that the IR-finite part of the amplitudes
also exponentiates.
Of course the separation between IR-divergent and IR-finite terms
is not unique since one could always add a constant to 
one 
while subtracting the same constant from the other.
Knowledge of the four-point amplitude alone does not give
enough information to resolve this ambiguity in a natural way, 
but after computing the five-point amplitude,
we will be able to determine a unique way of writing 
the universal IR-divergent part of the Higgs-regulated
scattering amplitude for any $n$.

\begin{figure}
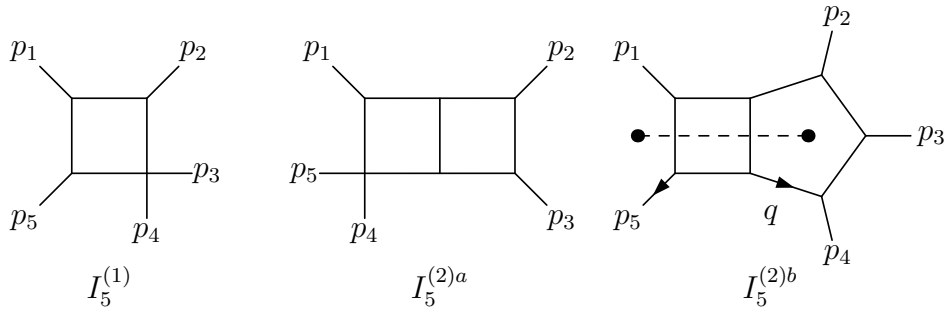

\begin{center}
\begin{feynartspicture}(375,125)(3,1)
\FADiagram{}
\FAProp(7.5,10.)(12.5,10.)(0.,){/Straight}{0}
\FAProp(7.5,15.)(12.5,15.)(0.,){/Straight}{0}
\FAProp(7.5,10.)(7.5,15.)(0.,){/Straight}{0}
\FAProp(12.5,10.)(12.5,15.)(0.,){/Straight}{0}
\FAProp(7.5,15.)(5.37878,17.1213)(0.,){/Straight}{0}
\FAProp(12.5,15.)(14.6213,17.1213)(0.,){/Straight}{0}
\FAProp(12.5,10.)(15.5,10.)(0.,){/Straight}{0}
\FAProp(12.5,10.)(12.5,7.)(0.,){/Straight}{0}
\FAProp(7.5,10)(5.37878,7.8787)(0.,){/Straight}{0}
\FALabel(4.37878,18.1213)[]{$p_1$}
\FALabel(15.6213,18.1213)[]{$p_2$}
\FALabel(16.5,10.)[]{$p_3$}
\FALabel(12.5,6.)[]{$p_4$}
\FALabel(4.37878,6.8787)[]{$p_5$}
\FALabel(10.,2.5)[]{$I^{(1)}_5$}
\FADiagram{}
\FAProp(15.,10.)(17.1213,7.87868)(0.,){/Straight}{0}
\FAProp(5.,15.)(15.,15.)(0.,){/Straight}{0}
\FAProp(15.,10.)(5.,10.)(0.,){/Straight}{0}
\FAProp(5.,10.)(5.,15.)(0.,){/Straight}{0}
\FAProp(10.,10.)(10.,15.)(0.,){/Straight}{0}
\FAProp(15.,10.)(15.,15.)(0.,){/Straight}{0}
\FAProp(5.,15.)(2.87868,17.1213)(0.,){/Straight}{0}
\FAProp(15.,15.)(17.1213,17.1213)(0.,){/Straight}{0}
\FAProp(5.,10.)(5.,7.)(0.,){/Straight}{0}
\FAProp(5.,10.)(2.,10.)(0.,){/Straight}{0}
\FALabel(1.,10.)[]{$p_5$}
\FALabel(5.,6.)[]{$p_4$}
\FALabel(1.87868,18.1213)[]{$p_1$}
\FALabel(18.1213,18.1213)[]{$p_2$}
\FALabel(18.1213,6.8787)[]{$p_3$}
\FALabel(10.,2.5)[]{$I_5^{(2)a}$}
\FADiagram{}
\FAProp(3.6529,15.)(8.6529,15.)(0.,){/Straight}{0}
\FAProp(8.6529,15.)(8.6529,10.)(0.,){/Straight}{0}
\FAProp(8.6529,10.)(3.6529,10.)(0.,){/Straight}{0}
\FAProp(3.6529,10.)(3.6529,15.)(0.,){/Straight}{0}
\FAProp(8.6529,15.)(13.4082,16.5451)(0.,){/Straight}{0}
\FAProp(8.6529,10.)(13.4082,8.45492)(0.,){/Straight}{1}
\FALabel(10.0306,7.22746)[]{$q$}
\FAProp(13.4082,8.45492)(16.3471,12.5)(0.,){/Straight}{0}
\FAProp(13.4082,16.5451)(16.3471,12.5)(0.,){/Straight}{0}
\FAProp(3.6529,15.)(1.5316,17.1213)(0.,){/Straight}{0}
\FAProp(3.6529,10.)(1.5316,7.87868)(0.,){/Straight}{1}
\FAProp(16.3471,12.5)(19.3471,12.5)(0.,){/Straight}{0}
\FAProp(13.4082,8.45492)(14.1085,5.53781)(0.,){/Straight}{0}
\FAProp(13.4082,16.5451)(14.1085,19.4622)(0.,){/Straight}{0}
\FALabel(0.5316,6.87868)[]{$p_5$}
\FALabel(0.5316,18.1213)[]{$p_1$}
\FALabel(14.5085,4.53781)[]{$p_4$}
\FALabel(14.5085,20.4622)[]{$p_2$}
\FALabel(20.7611,12.5)[]{$p_3$}
\FALabel(10.,2.5)[]{$I_5^{(2)b}$}
\FAVert(12.5,12.5){0}
\FAProp(12.5,12.5)(1.1529,12.5)(0.,){/ScalarDash}{0}
\FAVert(1.1529,12.5){0}
\end{feynartspicture}
\end{center}
\caption{Dual conformal scalar integrals contributing to 
the five-particle amplitude at one and two loops.  
The dashed line indicates that the integral
contains the loop-momentum-dependent numerator factor $(q + p_5)^2+m^2$.
}
\label{fig-five-point}
\end{figure}

As in ref.~\cite{Henn:2010bk} we evade a Feynman diagram calculation
by beginning with the ansatz that 
Higgs-regulated five-point loop amplitudes 
(normalized, as usual, by dividing by the corresponding tree amplitudes)
can be expressed as linear combinations of all possible dual conformally
invariant scalar integrals.
At one loop there is a unique integral $I_5^{(1)}(s_i)$
(see fig.~\ref{fig-five-point})
that can appear in the ansatz
\be
\label{unique}
M_5^{(1)} = - \frac{1}{4} \sum_{\rm cyclic} s_1 s_5 I_5^{(1)}(s_i)
+ \cO(m^2),
\ee
where
$+ \cO(m^2)$ stands for potential parity-odd terms (see below),
\be
s_i = (p_i + p_{i+1})^2, \qquad\qquad i=1, \ldots, 5
\ee
(with $p_6 \equiv p_1$),
and the sum in \eqn{unique} runs over the five cyclic 
permutations of the external momenta $p_i$.
The small $m^2$ expansion of $I_5^{(1)}(s_i)$
is given in appendix \ref{app-five-point}.
After summing over cyclic permutations, the one-loop
amplitude simplifies to 
\be
\label{fivepointoneloop} 
M_5^{(1)} = \sum_{i=1}^5 \left[- \frac{1}{4} 
\log^2 \left(\frac{s_i}{m^2}\right) \right] + F_5^\One
+ \cO(m^2)
\ee
where the corresponding finite remainder $F_5^{(1)}$ is given by\footnote{
Recently it has been shown \cite{Hodges:2010kq,Mason:2010pg}
that, when expressed in momentum-twistor
variables, the expression in
eqs.~(\ref{fivepointoneloop}) and~(\ref{oneloopF5})
computes the volume of a 4-simplex in AdS${}_5$ with 5
(regulated) points on the boundary.}
\be
\label{oneloopF5}
F_5^{(1)} = - \frac{1}{4} \sum_{i=1}^5
\left[
\log \left( \frac{s_{i}}{s_{i+1}}\right)
\log\left(\frac{s_{i-1}}{s_{i+2}}\right) -
\frac{\pi^2}{3}
\right]
\ee
with $s_{i+5} = s_i$.

At two loops there are two
dual conformal invariant scalar integrals
that contribute to the amplitude:
the double box $I^{(2)a}_5 (s_i)$
and the pentagon-box $I^{(2)b}_5 (s_i)$
with a numerator factor involving the pentagon loop momentum
(see fig.~\ref{fig-five-point}).
The coefficients of these integrals in
the amplitude $M_5^{(2)}(s_i)$ are determined by the consistency
of infrared singularities,
leading to the ansatz
\be
M_5^{(2)} = - \frac{1}{8} \sum_{\rm cyclic}
\left[
s_1 s_2^2 I^{(2)a}_5(s_i) + s_3^2 s_4 I^{(2)a}_5{(s_{6-i})} + s_2 s_3 s_5 I^{(2)b}_5(s_i)
\right] + \cO(m^2),
\label{unique2}
\ee
where,
as in the four-point calculation,
the relative numerical coefficients of each diagram are precisely
the same as in dimensional regularization.

Out of abundance of caution we have included
$+ \cO(m^2)$
in both eqs.~(\ref{unique2})
and (\ref{unique})
in order to encapsulate possible parity-odd terms.
In dimensional regularization,
$M_5^\Two$ has a non-vanishing parity-odd contribution
starting at $\cO(\epsilon^{-1})$.
The parity-odd part of $\log M_5$, however, is $\cO(\epsilon)$
at one- and two-loop order \cite{Cachazo:2006tj, Bern:2006vw,DelDuca:2009ae} 
due to the non-trivial cancellation of
parity-odd terms in $M_{5}^{(2)}$ 
and $-{1 \over 2} (M_{5}^{(1)})^2$,
which also has a contribution starting at $\cO(\epsilon^{-1})$.
The vanishing of the parity-odd contribution to the Higgs-regulated
amplitude $M_5^\Two$,
together with the fact that terms that vanish
as $m \to 0 $ are not required to compute $\log M_{5}$ in 
Higgs regularization, 
implies that the parity-odd contribution to $\log M_{5}$ is
$\cO(m^2)$
at one and two loops.

The small $m^2$ expansions of 
$I^{(2)a}_5 (s_i)$ and $I^{(2)b}_5 (s_i)$
are given in appendix \ref{app-five-point}.
After summing over cyclic permutations, we find
\be
M_5^{(2)} = \frac{1}{2} \left( M_5^{(1)} \right)^2 
+ \sum_{i=1}^5 
\left[ \frac{\zeta_2}{4} \log^2\left(\frac{s_i}{m^2} \right) 
+ \frac{\zeta_3}{2} \log \left( \frac{s_i}{m^2} \right)
- \frac{\zeta_2}{4} \log^2 \left( \frac{s_i}{s_{i+1}} \right) 
+ \frac{\zeta_2}{4} \log^2 \left( \frac{s_i}{s_{i+2}}\right) 
- \zeta_4 \right]
+ \cO(m^2).
\ee
Using the identity
\be
\sum_{i=1}^5 
\log \left(s_i \over s_{i+3} \right) 
\log \left(s_{i+1} \over s_{i+2} \right) 
=
\sum_{i=1}^5 
\left[ 2\log \left(s_i \over m^2 \right) \log \left(s_{i+1} \over m^2 \right) 
     - 2\log \left(s_i \over m^2 \right) \log \left(s_{i+2} \over m^2 \right) 
    \right] 
\ee
we may rewrite this as
\be
\label{fivepointtwoloop} 
M_5^{(2)}- \frac{1}{2} \left( M_5^{(1)} \right)^2 
=
\sum_{i=1}^5 
\left[
\frac{\zeta_2}{4} 
\log^2\left(\frac{s_i}{m^2} \right) 
+ \frac{\zeta_3}{2} \log \left( \frac{s_i}{m^2} \right)
\right]
 - \zeta_2  \, F_5^\One + 
\tc_5^\Two 
+ \cO(m^2)
\ee
where $ \tc_5^\Two  = {5 \over 4} \zeta_4 $
and $F_5^\One (s_i)$ is given in \eqn{oneloopF5}.
\Eqn{fivepointtwoloop} is analogous to 
the two-loop five-point iterative relation
of refs.~\cite{Cachazo:2006tj, Bern:2006vw}.

It is instructive to rewrite the one- and two-loop four-point
amplitudes in a form similar to \eqns{fivepointoneloop}{fivepointtwoloop}.
Using \eqn{all-loop-higgs}, 
with $s = s_1 = s_3$ and $t = s_2 = s_4$ for four-particle kinematics,  
we have
\ba
\label{fourpointoneloop} 
M_4^{(1)}
&=& 
\sum_{i=1}^4 \left[- \frac{1}{4} 
\log^2 \left(\frac{s_i}{m^2}\right) \right] + F_4^\One
+ \cO(m^2),
\\
\label{fourpointtwoloop} 
M_4^{(2)} - \frac{1}{2} \left( M_4^{(1)} \right)^2 
&=&
\sum_{i=1}^4 
\left[
\frac{\zeta_2}{4} 
\log^2\left(\frac{s_i}{m^2} \right) 
+ \frac{\zeta_3}{2} \log \left( \frac{s_i}{m^2} \right)
\right]
 - \zeta_2  \, F_4^\One + 
\tc_4^\Two 
\ea
where $ \tc_4^\Two  = {3 \over 4} \zeta_4$ and 
\be
\label{oneloopF4}
F^{(1)}_4 = \frac{1}{8} \sum_{i=1}^4
\left[
\log^2\left(\frac{s_{i}}{s_{i+1}}
\right) + \pi^2\right] .
\ee
By re-expressing the constants in \eqns{fivepointtwoloop}{fourpointtwoloop} as
\be
\tc_n^\Two = {(2n-5) \over 4} \zeta_4  
\ee
we may combine \eqns{fivepointtwoloop}{fourpointtwoloop} into
\be
M_n^{(2)}- \frac{1}{2} \left( M_n^{(1)} \right)^2 
=
\sum_{i=1}^n 
\left[
\frac{\zeta_2}{4} 
\log^2\left(\frac{s_i}{m^2} \right) 
+ \frac{\zeta_3}{2} \log \left( \frac{s_i}{m^2} \right)
+ \frac{\zeta_4}{2}
\right]
 - \zeta_2  \, F_n^\One 
- {5 \over 4} \zeta_4
+ \cO(m^2),
\qquad n=4, 5.
\ee
Note that the common infrared divergence term
\be
\sum_{i=1}^n\left[ \frac{\zeta_2}{4}  \log^2\left(\frac{s_i}{m^2}\right) +
\frac{\zeta_3}{2}  \log\left(\frac{s_i}{m^2}\right)
+ \frac{\zeta_4}{2}
\right]
\ee
resembles the form expected from dimensional
regularization~\cite{Catani:1998bh,Sterman:2002qn,Anastasiou:2003kj}:
\be
-\left[ \zeta_2  + \zeta_3  \epsilon + \zeta_4  \epsilon^2 \right]
M^{(1)}_n(2 \epsilon).
\ee
It would be interesting to understand more precisely the relation
between these two forms of the infrared divergences, perhaps along
the lines of ref.~\cite{Mitov:2006xs}.\footnote{We are grateful to S.~Moch for
discussion and correspondence on this question.}

Armed with the above ingredients we are now in a position to
pose the $n$-point generalization of \eqn{all-loop-higgs} as
\be
\label{BDSnparticle}
\log M_n = \sum_{i=1}^n
\left[ - \frac{\gamma(a)}{16} 
\log^2 \left( \frac{s_i}{m^2} \right)
- \frac{\tilde{\cal G}_0(a)} {2}
\log\left( \frac{s_i}{m^2} \right)
+ \tf (a)
\right]
+ \frac{1}{4} \gamma(a) \, F_n^{(1)} + \cR_n + \tC (a)
+ \cO(m^2)
\ee
where
\be
\label{Cdef}
\tf(a)  =  \frac{\zeta_4}{2} a^2 + {\cal O}(a^3),
\qquad\qquad 
\tC(a) = - \frac{5 \zeta_4}{4} a^2 + {\cal O}(a^3)
\ee
and the remainder function $\cR_n$ vanishes for $n=4$ and $n=5$.

We can then proceed by defining the Higgs regularization analogs 
of the IR-finite functions $F_n$ introduced in ref.~\cite{Bern:2005iz} 
by subtracting the universal infrared singularities from the 
logarithm of the amplitude
\be
F_n =\lim_{m^2 \to 0} \left(  \log M_n 
-  \sum_{i=1}^n  \left[ - \frac{\gamma(a)}{16} 
\log^2 \left( \frac{s_i}{m^2} \right)
- \frac{\tilde{\cal G}_0(a)} {2}
\log\left( \frac{s_i}{m^2} \right)
+ \tf (a)
\right]
\right).
\ee
The precise forms of the one-loop functions 
(\ref{oneloopF5}) and (\ref{oneloopF4})
and the two-loop functions 
\ba
\label{finiteremainders}
F^{(2)}_4 &=& \sum_{i=1}^4\left[  - \frac{\zeta_2}{8}
\log^2\left( \frac{s_{i}}{s_{i+1}} \right)
- \frac{35 \zeta_4}{16}
\right],
\nonumber\\
F_5^{(2)} &=&
\sum_{i=1}^5
\left[
- \frac{\zeta_2}{4} \log^2\left(\frac{s_{i}}{s_{i+1}}\right)
+ \frac{\zeta_2}{4} \log^2\left(\frac{s_{i}}{s_{i+2}}\right)
- \frac{3\zeta_4}{2}
\right]
\ea 
differ (by additive constants)
from the corresponding expressions in dimensional regularization,
which may be found in ref.~\cite{Bern:2005iz}.
Through two loops, however,  the $n=4$ and $n=5$ IR-finite functions
satisfy a similar iterative relation
\begin{equation}
F_n = \frac{1}{4} \gamma(a) F_n^{(1)} + \tC(a) + {\cal O}(a^3) , 
\qquad n=4, 5 \,.
\end{equation}
While the constants $ \tilde{\cG}_{0}(a)$ and $\tc_n(a)$  
defined above take different values in Higgs and in dimensional regularization,
it is fascinating to note that the two-loop value of $\tC(a)$ 
in \eqn{BDSnparticle} is identical to the corresponding value in dimensional
regularization \cite{Anastasiou:2003kj}.
Perhaps this is a coincidence, 
or perhaps it has a deeper explanation,
especially in light of the fact that the same value appears yet again in the
finite part of the two-loop lightlike polygon Wilson loop
after appropriate subtraction of UV divergences~\cite{Anastasiou:2009kna}.

For $n\ge 6$, we expect that
the ``remainder'' function $\cR_n$ in \eqn{BDSnparticle} is non-trivial,
just as in the corresponding formula
in dimensional regularization
\cite{Bern:2008ap,Drummond:2008aq}.
However it is natural to expect $\cR_n$ 
to take the same value in Higgs regularization as 
it does in dimensional regularization.
This is because the remainder function is an infrared-finite, 
dual conformally invariant quantity 
(as required by the dual conformal Ward identity), 
so it constitutes a good ``observable'' of SYM theory.
The 6-particle remainder function at two loops was first computed
numerically in ref.~\cite{Bern:2008ap}, 
where agreement with the corresponding remainder
function for lightlike hexagon Wilson 
loops \cite{Drummond:2007bm,Drummond:2008aq}
was established, and
an analytic expression has been given more recently
in refs.~\cite{DelDuca:2009au,DelDuca:2010zg,Zhang:2010tr}.

\section{Regge limits}
\setcounter{equation}{0}

In this section we discuss a number of features of the Regge limits of
Higgs-regulated amplitudes for $n \ge 5$.
As emphasized in footnote \ref{fn} such
limits may be taken in two different orders: limits (a) where all $m_i^2$
are first taken to be much smaller than all kinematical invariants,
and subsequently a Regge limit is taken, and limits (b) where the Regge
limits are taken first with various fixed kinematic invariants and fixed
masses $m_i^2$, which are subsequently taken to be much smaller than the
fixed kinematic invariants.  The Regge behavior of individual diagrams
can depend on the order, (a) or (b), in which these limits are taken.

\subsection{The five-point amplitude}

For the purposes of discussing the Regge limits of 
the five-gluon amplitude, we adopt the following 
parameterization of the kinematical invariants \cite{Brower:2008nm}
\ba
\label{fiveptparam}
s &=& (p_1+p_2)^2,\qquad
t_1 = (p_2 + p_3)^2,\qquad
s_1 = (p_3 + p_4)^2,\nonumber\\
s_2 &=& (p_4 + p_5)^2,\qquad
t_2 = (p_5 + p_1)^2, \qquad
s = s_1 s_2 /\kappa.
\ea
We will consider two different limits
of these invariants, as defined in ref.~\cite{Brower:2008nm}:
\begin{align}
\label{singleregge}
{\it Single \; Regge \; limit:  }& \quad \quad  
s \to \infty, \qquad s_1 \to \infty,
\quad\qquad\qquad \kappa, s_2, t_1, t_2 {\rm ~ fixed}\,,
\\
\label{doubleregge}
{\it Double \; Regge \;  limit:  }&  \quad  \quad 
s \to \infty, \qquad s_1 \to \infty, s_2 \to \infty
\qquad \kappa, t_1, t_2 {\rm ~ fixed}\,.
\end{align}
In the following two sections, we will consider
the Regge limits (a) and (b) for the five-point amplitude.
We will see that, at least through two loops, 
the amplitude is the same in both limits.

\subsection{Regge (a) limits for $n=5$}

If we wish to take Regge limit (a) of the five-gluon amplitude,
we can start with the conjectured ansatz (\ref{BDSnparticle}),
in which the small $m^2$ limit has already been taken.
In terms of the parameters (\ref{fiveptparam}),
\eqn{BDSnparticle} for $n=5$ takes the form 
\ba
\label{fivepointregge}
\log M_5
&=& 
\omega(t_1)
\log \left(s_1 \over m^2\right)
+\omega(t_2)
\log \left(s_2 \over m^2\right)
\nonumber\\
&+& {1 \over 16} \gamma(a) \left[ 
- \log^2 \left(t_1 \over t_2\right)
-\log^2 \left(\kappa\over m^2\right)
+2 \log \left(\kappa\over m^2\right)
\log \left( t_1 t_2 \over m^4\right)
\right]
\nonumber\\
&-& {1 \over 2} \tilde{\cG}_{0} \left[ 
\log \left(t_1 \over m^2 \right) 
+\log \left(t_2 \over m^2 \right) 
-\log \left(\kappa\over m^2\right) \right]
+\cO(m^2)
\ea
where
\be
\label{reggetraj}
\omega(t)= \alpha(t)-1 
=
-  {1 \over 4} \gamma(a) 
\log \Bigl( \toverm \Bigr)
- \tilde{\cG}_{0} (a)
\ee
is the same trajectory as in the four-point function (\ref{intro-alpha}).
\Eqn{fivepointregge} is equivalent to \cite{Brower:2008nm,Bartels:2008ce}
\be
\label{brower}
M_5 = \left( s_1 \over m^2 \right)^{\omega(t_1)}
      \left( s_2 \over m^2 \right)^{\omega(t_2)}
	F(t_1, t_2, \kappa)\,,
\ee
which exhibits the expected factorization.  From \eqn{brower} it 
is straightforward to take
the Regge limit to obtain the following expressions, 
separating the Regge behavior from the fixed term:
\begin{align}
\label{singlereggeamp}
{\it Single \; Regge \; (a)\; limit:  }& \quad \quad  M_5 \longrightarrow  \left( s_1 \over m^2 \right)^{\omega(t_1)} \left[
      \left( s_2 \over m^2 \right)^{\omega(t_2)} 
	F(t_1, t_2, \kappa) \right] \,, \\
\label{doublereggeamp}
{\it Double \; Regge \; (a)\; limit:  }&  \quad  \quad M_5 \longrightarrow  \left( s_1 \over m^2 \right)^{\omega(t_1)} 
      \left( s_2 \over m^2 \right)^{\omega(t_2)} 
\left[	F(t_1, t_2, \kappa) \right] \,.
\end{align}

Equations (\ref{singlereggeamp}) and (\ref{doublereggeamp}) 
have the same form as in
ref.~\cite{Brower:2008nm}, but with a different
value for the constant term in the
trajectory function (\ref{reggetraj}),
and a different $F(t_{1},t_{2},\kappa)$ due to the difference 
between \eqn{fivepointregge} and the analogous BDS result.

\subsection{Regge (b) limits for $n=5$}

To obtain the Regge (b) limits of the five-point amplitude,
one must start with scalar integrals 
with finite $m$, take the Regge limit first,
and only afterwards take $m$ small.  
We have evaluated the five-point one- and two-loop diagrams in this way
and have obtained results identical to \eqn{singlereggeamp}
in the single Regge limit and to \eqn{doublereggeamp}
in the double Regge limit.
Hence, at least to two-loop order, 
the amplitude is independent\footnote{
In an ``unphysical'' Regge limit  $s \to \infty$
with $s_1, s_2, t_1, t_2$ fixed, 
described on p. 177 of ref.~\cite{Eden},
the one-loop amplitude does depend on the order of limits,
yielding $\log^2 s$ dependence in the Regge (a) limit, as 
can be seen from \eqn{fivepointregge}, 
but $\log s$ dependence in the Regge (b) limit.}
of the order in which the 
Regge limit is taken (even though the individual diagrams are not).

In ref.~\cite{Henn:2010bk},
an alternative approach to the Regge (b) limits was applied 
to explain the absence of double logarithms in the Regge limit
of the four-gluon amplitude.
We can do the same for various Regge (b) limits of five-gluon amplitudes.
Consider the theory at a different point
on the Coulomb branch, where the
scalar diagrams have internal lines with 
variable masses along the periphery
(and vanishing masses in the interior).
Let $m_i$ be the mass of the line(s) connecting
$p_{i-1}$ and $p_i$.
The masses of the external lines are given 
by $p_i^2 = - (m_i-m_{i+1})^2$.
Due to dual conformal symmetry,
the amplitude depends only on the 
dual conformal invariants
\be
u_{i,i+2} =\frac{ m_i m_{i+2}} { (p_i + p_{i+1})^2 + (m_i - m_{i+2})^2 } .
\ee
By making various choices for $m_i$, we can reproduce 
the single and double Regge limits.

\vfil\break

\begin{figure}[t]
\psfrag{x1}[cc][cc]{$x_{1}$}
\psfrag{x2}[cc][cc]{$x_{2}$}
\psfrag{x3}[cc][cc]{$x_{3}$}
\psfrag{x4}[cc][cc]{$x_{4}$}
\psfrag{x5}[cc][cc]{$x_{5}$}
\psfrag{p1}[cc][cc]{$p_{1}$}
\psfrag{p2}[cc][cc]{$p_{2}$}
\psfrag{p3}[cc][cc]{$p_{3}$}
\psfrag{p4}[cc][cc]{$p_{4}$}
\psfrag{p5}[cc][cc]{$p_{5}$}
\psfrag{A}[cc][cc]{(ii)}
\psfrag{B}[cc][cc]{(i)}
 \centerline{
 {\epsfxsize10cm  \epsfbox{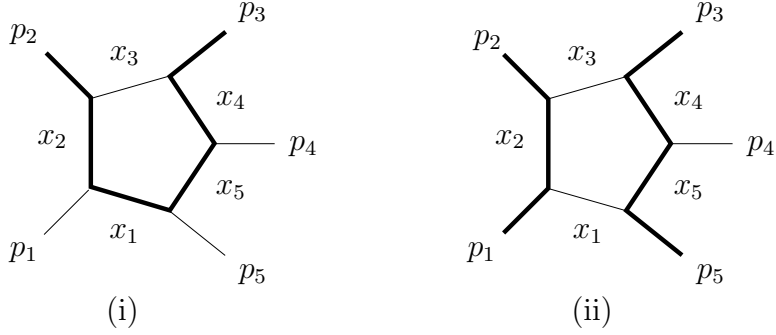}} 
}
\caption{\small
Two-mass configurations illustrating the 
(i) single and (ii) double Regge (b) limits for the five-point amplitude.
Fat and thin lines along the periphery denote particles of mass $M$ and $m$ 
respectively.
Fat and thin exterior lines denote particles of mass $M-m$ and 0 respectively.
}
\label{fig-twomass1}
\end{figure}

\bigskip\noindent{\it Single Regge (b) limit:  }

By making the choice $m_3=m$, and $m_i = M$ for $i \neq 3$,  we have
\be 
u_{13}=  \frac{Mm}{s + (M-m)^2}, \qquad
u_{24} = \frac{M^2}{t_1}, \qquad
u_{35} = \frac{Mm}{s_1 + (M-m)^2}, \qquad
u_{41} = \frac{M^2}{s_2}, \qquad
u_{52} = \frac{M^2}{t_1} \,.
\ee
Then taking the limit $m \ll M$ yields 
$u_{13}, u_{35} \ll u_{24}, u_{41},  u_{52}$,
which is equivalent to the single Regge limit (\ref{singleregge}).
The resulting diagrams (e.g., see fig.~\ref{fig-twomass1}(i)) 
cannot have collinear divergences
because the massless external lines never connect to the
light mass internal lines.
The amplitude therefore has at most a simple $\log m$ IR divergence,
which corresponds to $\log u_{35}$ (or $\log u_{13} $)
and therefore to a simple $\log s_1$  dependence in the 
single Regge limit, in agreement with \eqn{fivepointregge}.

\bigskip\noindent{\it Double  Regge (b) limit:  }

By choosing $m_1= m_3=m$, and $m_2 = m_4 = m_5 = M$, we have
\be 
u_{13} = \frac{m^2}{s}, \qquad
u_{24} = \frac{M^2}{t_1}, \qquad
u_{35} = \frac{Mm}{s_1 + (M-m)^2}, \qquad
u_{41} = \frac{Mm}{s_2 + (M-m)^2}, \qquad
u_{52} = \frac{M^2}{t_1} \,.
\ee
Taking the limit $m \ll M$
yields $u_{13} \ll u_{35}, u_{41}  \ll u_{24}, u_{52}$, 
which is equivalent to the double Regge limit (\ref{doubleregge}).
Again, none of the diagrams that contribute to the amplitude
in this limit has collinear divergences (e.g., see fig.~\ref{fig-twomass1}(ii)),
hence the amplitude has at most a simple $\log m$ IR divergence,
which corresponds to $\log u_{35}$ or $\log u_{41}$,
and therefore to a simple $\log s_1 $ or $\log s_2 $ dependence 
in the double Regge limit, in agreement with \eqn{fivepointregge}.

\subsection{Regge (a) limits for $n \ge 6$}
\label{regge-a-6}

To consider various Regge (a) limits of $n \ge 6$ amplitudes, 
we may start with the conjectured ansatz (\ref{BDSnparticle}),
in which the small $m^2$ limit has already been taken.

For $n=4$ and $n=5$ we have seen that $F^\One_n$,  
the IR-finite part of the one-loop amplitude,
has the same dependence on kinematic variables 
(up to an additive constant)
as the corresponding function in dimensional regularization.
Any difference in the form of $F^\One_n$ for $n \ge 6$
between Higgs-regulated amplitudes and dimensionally-regulated 
amplitudes should be dual conformal invariant,
and therefore a function of dual cross-ratios.
Any such function remains finite in Euclidean Regge 
limits \cite{Brower:2008nm}.
Similarly, the remainder function $\cR_n$ in either
Higgs regularization or in dimensional regularization
remains finite in Euclidean Regge limits,
as discussed in sec.~7 of ref.~\cite{Brower:2008nm}.
Therefore, the Regge (a) limits of Higgs-regulated amplitudes 
are equivalent to the Euclidean Regge limits 
of dimensionally-regulated amplitudes discussed 
in refs.~\cite{Brower:2008nm,Brower:2008ia},
but with the Regge trajectories given by \eqn{reggetraj}.
Similarly the Regge vertex functions will be analogous to those in 
ref.~\cite{Brower:2008nm}, but will differ in detail due 
to the difference in the IR-regulator scheme.

It is possible that $\cR_{n}$ could contribute to Regge limits in
the physical region, as these may involve contributions which do
not remain finite.  However, it is known that there are difficulties
in continuing $M_{n}$ ($n\ge 6$) from the Euclidean to the physical
region \cite{Bartels:2008ce,Bartels:2008sc,Brower:2008ia}.  These same
difficulties would be present for Higgs-regulated amplitudes, so we do
not consider these latter limits further.

In conclusion, there are no important differences between 
the Regge (a) limits of Higgs-regulated vs. dimensionally-regulated
$n$-gluon amplitudes.

\subsection{Regge limit (b) for $n = 6$}

We have not attempted to compute the amplitudes for $n\geq 6$ in any
of the Regge (b) limits defined in  \cite{Brower:2008nm,Brower:2008ia}
to ascertain whether they are equivalent to the Regge (a)
limits discussed in the previous subsection.

However, we would like to point out that the analysis of some of these
limits could be facilitated by considering Higgs-regulated amplitudes 
at various points on the Coulomb branch.
Let $m_{i}$ be the mass of the lines on the periphery of a diagram
connecting external lines $p_{i-1}$ and $p_{i}$, 
with vanishing masses in the interior. 
We consider points on the Coulomb branch involving just two
distinct mass assignments $m$ and $M$ (with $m \ll M$).

There are four inequivalent configurations (for six-point amplitudes)
that yield at most single logarithmic dependence on the 
IR-regulator mass $m$.
(These are configurations where 
no two {\it adjacent} $m_i$ are equal to $m$.)
We can classify these configurations according to the number of 
lines with the small mass $m$.

\bigskip\noindent{\it One line with small mass.} 

If we set $m_1 = m$ and $m_2$ through $m_6$ equal to $M$,
then none of the massless external lines are attached to the
line(s) on the periphery with mass $m$.
Hence there are no collinear divergences, and the 
amplitude only goes as $\log m$.

\bigskip\noindent{\it Two lines with small mass.} 

There are two possibilities involving two lines with small mass. 
The first one consists in setting
 $m_1 = m_3 = m$, with the remaining masses given by $M$.
The second consist in setting $m_1 = m_4 = m$,
with the remaining masses given by $M$.
The latter corresponds to the Mueller-Regge limit 
discussed in section 7.4 of ref.~\cite{Brower:2008ia}.

\bigskip\noindent{\it Three lines with small mass.} 

If the external masses alternate between $m$ and $M$
(i.e.  $m_1 = m_3 = m_5 = m$ and $m_2 = m_4 = m_6 =M$),
one obtains
the poly-Regge limit of section  7.3 of 
ref.~\cite{Brower:2008ia}.

\bigskip
No two-mass set-up has been found to give the single Regge limit or
helicity pole limit.

\section{Summary}
\setcounter{equation}{0}

In this paper we have continued the program of computing
higher-loop $\cN=4$ SYM planar $n$-gluon amplitudes 
and testing various conjectures using the Higgs 
regulator scheme proposed in ref.~\cite{Alday:2009zm} 
and further developed in ref.~\cite{Henn:2010bk}.
Specifically, we have extended the analysis to 
four loops for the four-gluon amplitude,
and to two loops for the five-gluon amplitude,
using Mellin-Barnes techniques to evaluate the integrals.
We have assumed that only scalar diagrams invariant under
extended dual conformal symmetry contribute to the amplitudes,
and with the same numerical coefficients as in dimensional regularization.
Although one has no {\it a priori} guarantee
that the set of diagrams contributing to the amplitude
in one IR-regulator scheme coincides with the set in another scheme,
all the results we obtained using this assumption are consistent with 
the universal IR-divergence structure of massless gauge theories
and also with the conjectured all-loop ansatz (\ref{intro-BDSnparticle}) 
for the IR-finite part.
For all cases considered, 
we have verified that the IR-finite parts
of the logarithm of the amplitudes have the same dependence on kinematic
variables as the corresponding functions in dimensionally-regulated
amplitudes (up to overall additive constants, which we determine).

We have also extended the study of the Regge behavior
of Higgs-regulated amplitudes which was begun in ref.~\cite{Henn:2010bk}.  
The Regge (a) limits
(in which the masses are first taken much smaller 
than all kinematic invariants, and then the Regge 
limit of kinematic variables is applied) 
can be understood by simply taking the 
kinematic limits of the ansatz (\ref{intro-BDSnparticle}),
in which $\cO(m^2)$ terms are already neglected.
Various (a) type limits are discussed for $n \ge 5$
which give essentially the same results as those 
from the dimensionally-regulated BDS ansatz, 
but with different expressions for the gluon trajectory 
and Regge vertices resulting from the different regulator scheme.

To study Regge (b) limits 
(in which the kinematical limits are first taken with fixed 
regulator masses, which are subsequently taken to be much smaller
than the fixed kinematic invariants)
one must evaluate the Regge limits of the individual diagrams
contributing to the amplitudes.
In the Regge (b) limit, certain classes of diagrams are dominant,
whereas in the Regge (a) limit no single class of diagrams dominates.
In ref.~\cite{Henn:2010bk},
it was shown that the leading-log approximation of the four-gluon
amplitude is dominated to all loop orders by the sum of vertical ladder
diagrams only.  In this paper, we showed that
the next-to-leading-log approximation
depends only on the vertical ladder diagrams, together with the class
of vertical ladder diagrams with a single H-shaped insertion.

One way of analyzing Regge (b) limits of diagrams
is to go to a different point on the Coulomb branch
involving several different masses.  
Examples are given for several Regge limits of $n=5$ and $n=6$ 
amplitudes using two different masses, 
although we have not been able to obtain all Regge (b) limits 
using two-mass configurations on the Coulomb branch.
Although the Regge (a) and Regge (b) limits of individual diagrams differ,
we have found that the full amplitudes are independent of the order of
limits in the cases that we have considered.

The results of this paper show that the Higgs regulator
for planar $\cN=4$ SYM amplitudes continues to
exhibit a number of practical and conceptual advantages compared to
other regulators,
the first signs of which were observed in 
ref.~\cite{Alday:2009zm,Henn:2010bk}.
On the practical side, the Higgs regulated multi-loop integrals we have
encountered so far have proven quite a bit simpler to evaluate than their
counterparts in dimensional regularization.  
One consequence of this is that we have been
able to compute the four-loop cusp anomalous dimension with numerical
precision
five orders of magnitude greater than ref.~\cite{Bern:2006ew} 
and two orders of magnitude greater than ref.~\cite{Cachazo:2006az}.
The crucial conceptual advantage of the Higgs regulator is that
it preserves the remarkable (extended) dual conformal symmetry 
which has recently played such an important role
in unlocking the hidden structure of SYM amplitudes.
While dimensional regularization seems completely at odds 
with modern (inherently four-dimensional) twistor-space methods,
it is greatly encouraging that the Higgs regulator can be very naturally
implemented in momentum twistor space, 
as seen for example in the beautiful recent
results of ref.~\cite{Hodges:2010kq,Mason:2010pg},
which we suspect are just the tip of the iceberg.

\section*{Acknowledgments}

It is a pleasure to thank S.~Moch, C.-I.~Tan and A.~Volovich for discussions
and correspondence.  J.H. is grateful to Brown University
and to the Institute for Advanced Study, where part of this
work was done, for hospitality.

\vfil\break

\appendix

\section{Regge limits of four-loop four-point integrals}
\setcounter{equation}{0}
\label{app-four-loop} 
In this appendix, we list the results for the 
Regge (b) limit of the each of the four-loop integrals 
that contribute to the four-loop amplitude.
For compactness, we employ the notation
$\{ a_{1} , \ldots , a_{n} \} \equiv  
\sum_{m=1}^{n} a_{m} \log^{n-m} v + \cO(v)$ .
\begin{align}
 I_{4a}(s,t) = &
    \log u \, \{ \frac{8}{315} ,0,  \frac{8}{45} \pi^2 , 0 , \frac{56}{135} \pi^4,  0 , 252.30\ldots, {\rm const} \} + \cO(\log^0 u) \,. \\
 I_{4a}(t,s) =&  \log^4 u \, \{ \frac{2}{3} , 0,0,0 \} \\
 & + \log^3 u \, \{  -\frac{8}{3} , 0 , -\frac{8}{3} \pi^2 , - 8 \zeta_{3} ,0,0 \} \nonumber  \\
 & + \log^2 u \, \{ \frac{214}{45} ,0, \frac{100}{9} \pi^2 ,  \frac{64}{3}  \zeta_{3} ,  \frac{154}{45} \pi^4 , 183.07\ldots, 11.55\ldots \} \nonumber \\
 & +   \log u \, \{ -\frac{1352}{315} ,0, - \frac{736}{45} \pi^2 , - \frac{68}{3} \zeta_{3} ,  -\frac{608}{45} \pi^4, -580.84\ldots , -1536.93\ldots,  {\rm const} \} + \cO(\log^0 u).  \nonumber \\
 I_{4b}(s,t) =&  \log^2 u \, \{ \frac{4}{15} ,0, \frac{8}{9} \pi^2 , 0 ,  \frac{28}{45} \pi^4 , 0, 0 \}  \\
 & +   \log u \, \{ -\frac{257}{630} ,0, - \frac{101}{45} \pi^2 , - \frac{16}{3} \zeta_{3} , -\frac{424}{135} \pi^4, -88.16\ldots, -940.02\ldots, {\rm const} \} + \cO(\log^0 u) \,. \nonumber \\
  I_{4b}(t,s) =&
    \log u \, \{ \frac{8}{63} ,0,  \frac{16}{45} \pi^2 , 0 , \frac{8}{27} \pi^4,  0 , 65.11\ldots, {\rm const} \} + \cO(\log^0 u) \,. \\
 I_{4c}(s,t) =&
    \log u \, \{ \frac{16}{315} ,0,  \frac{8}{45} \pi^2 , 0 , \frac{32}{135} \pi^4,  0 , 105.80\ldots, {\rm const} \} + \cO(\log^0 u) \,. \\
 I_{4c}(t,s) =&   \log^3 u \, \{  \frac{8}{9} , 0 , \frac{8}{9} \pi^2 , 0 ,0,0 \} \\
 & + \log^2 u \, \{ - \frac{121}{45} ,0, -\frac{56}{9} \pi^2 , - \frac{32}{3}  \zeta_{3} , - \frac{32}{15} \pi^4 , -56.52\ldots, 0 \} \nonumber \\
 & +   \log u \, \{ \frac{1963}{630} ,0,  \frac{539}{45} \pi^2 ,  \frac{62}{3} \zeta_{3} , \frac{1478}{135} \pi^4, 622.32\ldots , 1619.96\ldots,  {\rm const} \} + \cO(\log^0 u) \,. \nonumber\\
 I_{4d}(s,t) =&
    \log u \, \{ \frac{8}{315} ,0,  \frac{8}{45} \pi^2 , 0 , \frac{56}{135} \pi^4,  0 , 252.30\ldots, {\rm const} \} + \cO(\log^0 u) \,. 
\\
 I_{4d}(t,s) =&   \log^3 u \, \{  \frac{8}{9} , 0 , \frac{8}{9} \pi^2 , 0 ,0,0 \} \\
 & + \log^2 u \, \{ - \frac{44}{15} ,0, -\frac{52}{9} \pi^2 , - 8  \zeta_{3} , - \frac{22}{9} \pi^4 , -49.77\ldots, 0 \} \nonumber \\
 & +   \log u \, \{ \frac{1132}{315} ,0,  12 \pi^2 ,  \frac{56}{3} \zeta_{3} , \frac{428}{45} \pi^4, -191.36\ldots , 863.63\ldots,  {\rm const} \} + \cO(\log^0 u) \,. \nonumber
\\
 I_{4e}(s,t) =&
    \log u \, \{ \frac{8}{105} ,0,  \frac{16}{45} \pi^2 , 0 , \frac{56}{135} \pi^4,  0 , 130.22\ldots, {\rm const} \} + \cO(\log^0 u) \,. \\
 I_{4e}(t,s) =&    \log^2 u \, \{  \frac{8}{15} ,0, \frac{8}{9} \pi^2 , 0,  \frac{16}{45} \pi^4 , 0, 0 \} \\
 & +   \log u \, \{ - \frac{223}{210} ,0,  -\frac{143}{45} \pi^2 , - \frac{16}{3} \zeta_{3} , -\frac{22}{9} \pi^4, -49.77\ldots , -258.40\ldots,  {\rm const} \} + \cO(\log^0 u) \,. \nonumber
 %
\end{align}
\begin{align}
 I_{4f}(s,t) =&
    \log u \, \{ \frac{16}{45} ,0,  \frac{32}{45} \pi^2 , -\frac{4}{3} \zeta_{3} , \frac{2}{5} \pi^4,  -39.93 \ldots , 41.35\ldots, {\rm const} \} + \cO(\log^0 u) \,. \\
 I_{4f}(t,s) =&    \log^2 u \, \{  \frac{8}{9} ,0, \frac{16}{9} \pi^2 , 0,  \frac{8}{9} \pi^4 , 0, 0 \} \\
 & +   \log u \, \{ - \frac{484}{315} ,0,  -\frac{268}{45} \pi^2 , - 12 \zeta_{3} , -\frac{734}{135} \pi^4, -103.20\ldots , -907.47\ldots,  {\rm const} \} + \cO(\log^0 u) \,. \nonumber\\
 I_{4d_{2}}(s,t) =&
    \log u \, \{ 0 ,0, 0 , -\frac{4}{3} \zeta_{3} , \frac{2}{45} \pi^4,  -39.93\ldots , 41.34\ldots, {\rm const} \} + \cO(\log^0 u) \,. \\
 I_{4d_{2}}(t,s) =&
    \log u \, \{ 0 ,0, 0 , -\frac{4}{3} \zeta_{3} , \frac{2}{15} \pi^4,  -9.84\ldots , 28.48\ldots, {\rm const} \} + \cO(\log^0 u) \,. \\
 I_{4f_{2}}(s,t) =&    \log^2 u \, \{  \frac{8}{9} ,0, \frac{16}{9} \pi^2 , 0,  \frac{8}{9} \pi^4 , 0, 0 \} \\
 & +   \log u \, \{ - \frac{124}{105} ,0,  -\frac{236}{45} \pi^2 , - 16 \zeta_{3} , -\frac{136}{27} \pi^4, -109.95\ldots , -935.95\ldots,  {\rm const} \} + \cO(\log^0 u) \,. \nonumber\\
 I_{4f_{2}}(t,s) =&    I_{4f_{2}}(s,t) \,.
 \end{align}

\vfil\break
\section{Regge limit of the $L$-loop ladder with $H$-insertion}
\setcounter{equation}{0}
\label{app-H}

In this appendix, we evaluate the leading log contribution
(in the limit of large $s$)
of the dual conformal invariant $L$-loop vertical ladder 
diagram with one H-shaped insertion,
$I_{LH}$, described in sec.~\ref{sect-regge-fourpoint}.

\begin{figure}[ht]
\psfrag{x1}[cc][cc]{$x_{1}$}
\psfrag{x2}[cc][cc]{$x_{2}$}
\psfrag{x3}[cc][cc]{$x_{3}$}
\psfrag{x4}[cc][cc]{$x_{4}$}
\psfrag{p1}[cc][cc]{}
\psfrag{p2}[cc][cc]{}
\psfrag{p3}[cc][cc]{}
\psfrag{p4}[cc][cc]{}
\psfrag{La}[cc][cc]{(i)}
\psfrag{ILH}[cc][cc]{(ii)}
\psfrag{dots}[cc][cc]{\dots}
\psfrag{xj}[cc][cc]{$x_{j}$}
\psfrag{xjm}[cc][cc]{$x_{j-1}$}
\psfrag{xjp}[cc][cc]{$x_{j+1}$}
 \centerline{
 {\epsfxsize10cm  \epsfbox{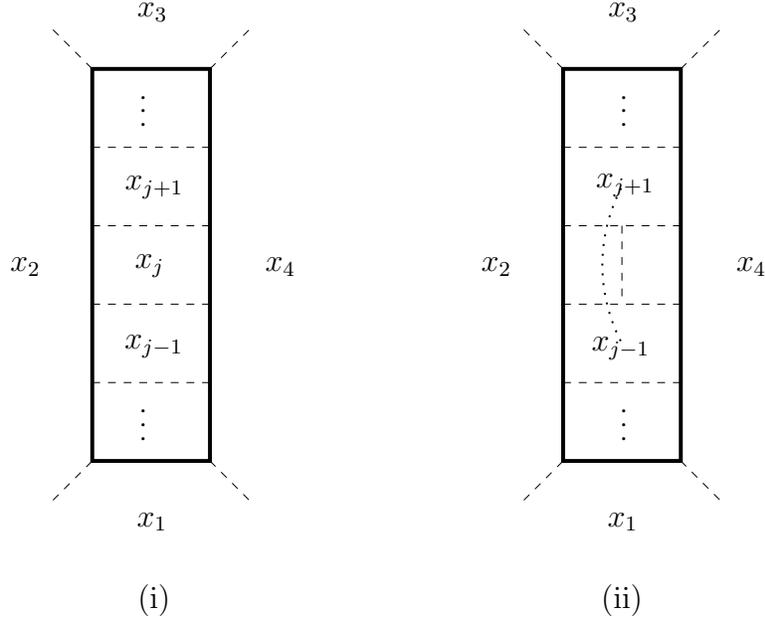}}
}
\caption{\small
(i) Vertical ladder diagram and (ii) vertical ladder with H-shaped
insertion (with numerator factor indicated by a dotted line) whose
leading log contributions are computed in this appendix.
}
\label{fig-appb}
\end{figure}

We begin by considering an $(L-1)$-loop vertical ladder 
diagram, with the external regions labeled by $x_1$ through $x_4$
(where $s = x_{13}^2$ and $t= x_{24}^2$),
and the loops labeled by $x_i$, with $i=5, \cdots, L+4$ 
(see fig.~\ref{fig-appb}(i)).
We replace the $j$th loop with a horizontal double loop, 
converting it into an $L$-loop diagram, as shown in fig.~\ref{fig-appb}(ii).
By dual conformal invariance, this diagram must be accompanied
by a factor of 
\be
\label{numer}
x^2_{13} \, x^{2(L-1)}_{24} \, x_{j-1,j+1}^2 .
\ee
We now perform the integration over the double box.
Its Feynman parameterization is 
\be
\label{doublebox}
\Idbox 
= 
\int_0^1 d\al_0 \,d\al_1 \,d\al_2 \,d\bet_1 \,d\bet_2 \,d\ga_1 \,d\ga_2 
\, {2A 
\,\,\delta( \al_0 +\al_1+\al_2 +\de_1 + \de_2 -1) 
\over (D + m^2 \sigma A)^3 }
\ee
where 
\ba
\label{param}
D &=& D_{24} t
     +D_{j-1,j+1} P_{j-1,j+1} 
     +D_{j-1,2} P_{j-1,2} 
     +D_{j+1,2} P_{j+1,2} 
     +D_{j-1,4} P_{j-1,4} 
     +D_{j+1,4} P_{j+1,4} 
\nonumber\\
A &=& \al_0 \al_1 + \al_0 \al_2 +\al_1 \al_2 +\al_0\de_1 + \al_1 (\de_1 + \de_2)+ \al_2 \de_2 + \de_1 \de_2 \nonumber\\ 
\sigma &=& \al_0 + \al_2 
\nonumber\\
\de_i 
&=&
 \bet_i + \ga_i
\ea
with $P_{ij} \equiv x_{ij}^2 \equiv (x_i - x_j)^2$.

The coefficients of $P_{ik}$ in \eqn{param} are given by
\ba
D_{24}  &=& \al_0 \al_1 \al_2 ,
\qquad  
D_{j-1,j+1} 
= 
( \al_2+\de_2) \bet_1 \ga_1 
+(\al_0+\de_1) \bet_2 \ga_2 
+ \al_1 (\bet_1+\bet_2) (\ga_1 + \ga_2),
\nonumber\\
D_{j-1,2} 
&=& 
\al_0 ( \al_1 \bet_1 + \al_2 \bet_1 + \al_1 \bet_2 + \de_2 \bet_1),
\qquad
D_{j+1,2} = \al_0 ( \al_1 \ga_1 + \al_2 \ga_1 + \al_1 \ga_2 + \de_2 \ga_1),
\nonumber\\
D_{j-1,4} 
&=& 
\al_2 ( \al_1 \bet_2 + \al_0 \bet_2 + \al_1 \bet_1 + \de_1 \bet_2),
\qquad
D_{j+1,4} 
= 
\al_2 ( \al_1 \ga_2 + \al_0 \ga_2 + \al_1 \ga_1 + \de_1 \ga_2)\,.
\nonumber\\
\ea
If the double box is inserted at one of the 
ends of the vertical ladder, 
the expression for $\sigma$ in \eqn{param}
will contain additional terms
$\bet_1 + \bet_2$ or $\ga_1 + \ga_2$,
but the results below will be unaffected by this change.

We assume, following chapter 8 of ref.~\cite{Gribov}, that the 
asymptotic behavior of multiloop integrals is dominated
by a region in the space of the loop momenta 
that can be described by a set of nested inequalities.
In our specific case, those inequalities imply that
the leading $s\to\infty$ behavior of $I_{LH}$
comes from the region where 
$P_{j-1,j+1}$ is large (i.e. a nonzero fraction of $s$)
with the other $P_{ik}$ appearing in \eqn{param} remaining finite.
This in turn implies that 
the main contribution to the subintegral (\ref{doublebox}) 
arises from the part of parameter space where 
$D_{j-1,j+1}$, the coefficient of $P_{j-1,j+1}$, is small.
This region may be identified by 
parameterizing \cite{Federbush}
\be
\bet_1 = \rho_1 \zeta_1, \qquad
\bet_2 = \rho_1 \bar\zeta_1, \qquad
\ga_1 = \rho_2 \zeta_2, \qquad
\ga_2 = \rho_2 \bar\zeta_2
\ee
where $\zeta_i$ runs from 0 to 1, and $\bar\zeta_i \equiv 1 - \zeta_i$.
In terms of these parameters, the main contribution to the subintegral
comes from the region where $\rho_1$ and $\rho_2$ are both small. 
Retaining only the lowest order terms  in $A$ and $D$,
\ba
A &=& \al_0 \al_1 + \al_0 \al_2 +\al_1 \al_2 + \cO(\rho_i),
\\
D_{24}  
&=&
\al_0 \al_1 \al_2 ,
\qquad
D_{j-1,j+1} 
= 
\rho_1 \rho_2 (\al_1 + \al_2 \zeta_1 \zeta_2 + \al_0 \bar\zeta_1 \bar\zeta_2) 
+ \cO(\rho^3) ,
\nonumber\\
D_{j-1,2} 
&=& 
\rho_1 \al_0 ( \al_1 + \al_2 \zeta_1) + \cO(\rho_1^2),
\qquad
D_{j+1,2} 
= 
\rho_2 \al_0 ( \al_1 + \al_2 \zeta_2) + \cO(\rho_2^2),
\nonumber\\
D_{j-1,4} 
&=& 
\rho_1 \al_2 (\al_1 + \al_0 \bar\zeta_1) + \cO(\rho_1^2),
\qquad
D_{j+1,4} 
 = 
 \rho_2 \al_2 (\al_1 + \al_0 \bar\zeta_2) + \cO(\rho_2^2)
\nonumber
\ea
we may approximate the term in the denominator of \eqn{doublebox} as
\be
\label{denom}
D + m^2 \sigma A \quad \approx \quad
\Gam_0 \rho_1 \rho_2 + \Gam_1 \rho_1 + \Gam_2 \rho_2 + \Gam_3 
\ee
with 
\ba
\Gam_0   &=& 
(\al_1 + \al_2 \zeta_1 \zeta_2 + \al_0 \bar\zeta_1 \bar\zeta_2)  P_{j-1,j+1}
\nonumber\\
\Gam_1   &=& 
\al_0 ( \al_1 + \al_2 \zeta_1) P_{j-1,2}
+\al_2 (\al_1 + \al_0 \bar\zeta_1) P_{j-1,4} 
\nonumber\\
\Gam_2   &=& 
\al_0 ( \al_1 + \al_2 \zeta_2) P_{j+1,2} 
+\al_2 (\al_1 + \al_0 \bar\zeta_2) P_{j+1,4}
\nonumber\\
\Gam_3   &=& 
\al_0 \al_1 \al_2 t  
+ (\al_0 \al_1 + \al_0 \al_2 +\al_1 \al_2 )(\al_0 + \al_2) m^2 \,.
\ea
In the large $s$ limit,  $\Gam_0 \gg \Gam_1 \sim \Gam_2 \sim \Gam_3$.
Inserting \eqn{denom} into \eqn{doublebox},
we may approximate the integral as \cite{Eden}
\be
\label{db}
\Idbox \approx \int_0^1 d\al_0 \,d\al_1\,d\al_2
\, \delta( \al_0 +\al_1+\al_2 -1) 
\, (\al_0 \al_1 + \al_0 \al_2 +\al_1 \al_2)
\int_0^1 d\zeta_1 d\zeta_2
\, J
\ee
with
\be
J = \int_0^{\eta_1} d\rho_1
\int_0^{\eta_2} d\rho_2
{2 \rho_1 \rho_2 
\over (\Gam_0 \rho_1 \rho_2 + \Gam_1 \rho_1 + \Gam_2 \rho_2 + \Gam_3 )^3 }
\ee
where the factors of $\rho_i$ in the numerator result 
from the Jacobian in the change of variables. 
To capture the dominant behavior of the integral, we only need
integrate $\rho_1$ and $\rho_2$ over a small region near the origin;
the exact values of the upper limits $\eta_1$, $\eta_2$ are unimportant 
but are both taken to be $\ll 1$ to justify the neglect of higher order terms 
in $\rho_i$ in the integrand and the dropping of $\delta_i$ 
from the argument of the Dirac delta function.
The integral over $\rho_i$ yields 
\be
\label{J}
J \approx {\Gam_3 \over (\Gam_0 \Gam_3 - \Gam_1 \Gam_2)^2 } 
\log 
\left(  (\Gam_0 \eta_1 \eta_2 + \Gam_1 \eta_1 + \Gam_2 \eta_2 + \Gam_3 )
\Gam_3 
\over
(\Gam_1 \eta_1+\Gam_3 ) (\Gam_2 \eta_2 + \Gam_3 )
\right)
\approx
{1 \over \Gam^2_0 \Gam_3 }
\log \Gam_0 
\ee
where in the last step we used the approximation
$\Gam_0 \gg \Gam_1 \sim \Gam_2 \sim \Gam_3$.
Inserting \eqn{J} into \eqn{db}, we find
\be
\label{integdouble}
\Idbox \approx 
{K'(v) \over t} \  { \log P_{j-1,j+1} \over P^2_{j-1,j+1} }
\ee
where $v = m^2/t$ and 
\be
K'(v) = \int_0^1 d\al_0 \,d\al_1 \,d\al_2
\int_0^1 d\zeta_1 d\zeta_2
{ \delta( \al_0 +\al_1+\al_2 -1) 
\,(\al_0 \al_1 + \al_0 \al_2 +\al_1 \al_2)
 \over 
(\al_1 + \al_2 \zeta_1 \zeta_2 + \al_0 \bar\zeta_1 \bar\zeta_2)^2
\left[
 \al_0 \al_1 \al_2   
+ (\al_0 \al_1 + \al_0 \al_2 +\al_1 \al_2 )(\al_0 + \al_2) v 
\right]
}.
\ee
Thus in the $s \to \infty$ limit, 
integrating out the double box is equivalent to inserting 
\eqn{integdouble} into the remaining integral.

We observe that $K'(v)$ may be identified as the coefficient of 
$(\log s)/(s^2 t)$ in the two-loop horizontal ladder diagram
in the asymptotic $s\to \infty$ regime.    When $v$ is small,
$K'(v)$ can be explicitly evaluated to give \cite{Henn:2010bk}
\be
K'(v) =  -\frac{4}{3} \log^3 v - \frac{4}{3} \pi^2 \log v + \cO(v) \,.
\ee

Next we consider an 
$(L-1)$-loop vertical ladder integral $I_{L-1,a}$,
supplemented with a numerator factor of
$x_{13} x^{L-1}_{24} $
to make it dual conformal invariant.
Consider just the subintegral over the $j$th loop
\be
\label{singlebox}
\Ibox 
=
\int_0^1 d\al_0 \,d\al_1 \,d\bet_1 \,d\ga_1 
{\delta( \al_0 +\al_1 +\bet_1 +\ga_1 -1) 
 \over (\Gam_0 \bet_1 \ga_1 + \Gam_1 \bet_1 + \Gam_2 \ga_1 + \Gam_3)^2 }
\ee
where
\ba
\Gam_0 &=&  P_{j-1,j+1} 
\nonumber\\
\Gam_1 &=&  \al_0 \,P_{j-1,2} + \al_1 \,P_{j-1,4}  
\nonumber\\
\Gam_2 &=&  \al_0 \,P_{j+1,2} + \al_1 \,P_{j+1,4} 
\nonumber\\
\Gam_3 &=&   \al_0 \al_1  t + m^2 (\al_0+\al_1) \,.
\ea
As before, the leading $s \to\infty$ behavior of the vertical ladder 
integral comes from the region of loop momentum space where
$P_{j-1,j+1}$ is much larger than the other $P_{ik}$, i.e.,
where 
$\Gam_0 \gg \Gam_1 \sim \Gam_2 \sim \Gam_3$.
This in turn implies that 
the main contribution to the subintegral (\ref{singlebox}) 
arises from the region of parameter space where 
$\bet_1$ and $\ga_1$ are both small.
We may therefore approximate \eqn{singlebox} as
\be
\Ibox \approx \int_0^1 d\al_0 \,d\al_1 
 \delta( \al_0 +\al_1 -1) 
\int_0^{\eta_1} d\bet_1 
\int_0^{\eta_2} d\ga_1 
{1 \over (\Gam_0 \bet_1 \ga_1 + \Gam_1 \bet_1 + \Gam_2 \ga_1 + \Gam_3)^2 } \,.
\ee
The integral over $\bet_1$ and $\ga_1$ yields
\be
{1 \over \Gam_0 \Gam_3 - \Gam_1 \Gam_2 } 
\log 
\left(  
(\Gam_0 \eta_1\eta_2 + \Gam_1 \eta_1 + \Gam_2 \eta_2 + \Gam_3) \Gam_3 
\over
(\Gam_1 \eta_1 + \Gam_3) (\Gam_2 \eta_2 + \Gam_3)
\right)
\approx {1 \over \Gam_0 \Gam_3} \log \Gam_0
\ee
where in the last step we have used the approximation 
$\Gam_0 \gg \Gam_1$, $\Gam_2$, $\Gam_3$.
Hence we obtain for the subintegral
\be
\label{integsingle}
\Ibox \approx {K(v) \over t} \ { \log P_{j-1,j+1} \over P_{j-1,j+1} } 
\qquad {\rm  where } \qquad 
K(v) 
= \int_0^1 d\al_0 \,d\al_1 
{ \delta( \al_0 +\al_1 -1)  \over 
\al_0 \al_1   + v}  \,.
\ee
Thus, in the $s \to \infty$ limit, 
integrating out a single box is equivalent 
to the insertion of \eqn{integsingle} in the remaining integral.
Here $K(v)$ is just the coefficient of  $(\log s)/(s t)$ 
in the one-loop box diagram in the asymptotic $s\to \infty$ regime.    
When $v$ is small,
it can be explicitly evaluated to give 
\be
K(v) =  -2 \log v + \cO(v) \,.
\ee

We note that \eqns{integdouble}{integsingle}
are quite similar, except that the powers of $P_{j-1,j+1}$
in the denominators differ.  This reflects the
fact that the two d-lines (described in sec.~\ref{sect-regge-fourpoint}) 
for the double box have 
length two whereas those for the single box only
have unit length.   Recall, however, that dual 
conformal invariance implies that the $I_{LH}$ diagram 
comes with a factor of $P_{j-1,j+1}$ in the numerator (\ref{numer}),
which cancels one of the factors in the denominator of \eqn{integdouble},
``promoting'' the d-lines to length one.
(This in turn raises the number of powers of $\log s$ in 
the asymptotic behavior of the integral by two.)

Taking into account all the numerator factors,
we see that the quotient of $I_{LH}$ and $I_{L-1,a}$ is
given by 
\be
{I_{LH} \over I_{L-1,a}}
={K'(v) \over K(v)} \,.
\ee
We know however from ref.~\cite{Eden,Henn:2010bk}  that
\be
\;\; \mathrel{\mathop{\rm lim}\limits_{u \ll v}}\quad
I_{L-1,a} =   {(-1)^{L-1} \over (L-1)!}  K(v)^{L-1}  \log^{L-1}  u
+ \cdots
\ee
from which we conclude that
\be
\;\; \mathrel{\mathop{\rm lim}\limits_{u \ll v}}\quad
I_{LH} 
= {(-1)^{L-1}  \over (L-1)!} \,  K(v)^{L-2} \, K'(v)  \log^{L-1} u + \cdots
\ee
which is used in the main body of the paper in \eqn{LH}.

\vfil\break
\section{Five-point integrals}
\setcounter{equation}{0}
\label{app-five-point}

In this appendix, we give small-$m^2$ expansions for the
Higgs-regulated one- and two-loop five-point integrals 
shown in fig.~\ref{fig-five-point}.

The one-loop box integral $I^{(1)}_5$
has the small $m^2$ expansion
\begin{multline}
 s_{1} s_{5} I^{(1)} =
\log^2 m^2 - 2 \log\left(s_{1} s_{5}\over s_{3}\right) \log m^2
\cr
+ 2 \log s_{1} \log s_{5} - \log^2 s_{3}
- 2\,\text{Li}_2\left(1 - \frac{s_{3}}{s_{1}}\right)
- 2\,\text{Li}_2\left(1 - \frac{s_{3}}{s_{5}}\right)
- \frac{\pi^2}{3}.
\end{multline}
Assembling everything into dimensionless ratios $s_{i}/m^2$, 
the result can be stated more succinctly as
\begin{equation}
s_{1} s_{5} I^{(1)}_5 \approx
2 \log\left( \frac{s_{1}}{m^2}\right)
\log \left(\frac{s_{5}}{m^2}\right) - \log^2 \left( \frac{s_{3}}{m^2}\right)
- 2\,\text{Li}_2\left(1 - \frac{s_{3}}{s_{1}}\right)
- 2\,\text{Li}_2\left(1 - \frac{s_{3}}{s_{5}}\right)
- \frac{\pi^2}{3}
+ \cO(m^2).
\end{equation}
The double-box integral $I^{(2)a}_5$ has the small $m^2$ expansion
\begin{equation}
\begin{split}
  s_{1} s_{2}^2 I^{(2) a} =&
- \frac{1}{4} \log^4 m^2
+ \log\left( s_{1} s_{2}\over s_{4}\right) \log^3 m^2
\cr
& + \left[
- \frac{3}{2} \log^2\left(s_{1} s_{2}\over s_{4}\right)
- 2\,\text{Li}_2\left(1 - \frac{s_{2}}{s_{4}}\right)
- \frac{\pi^2}{6}
\right] \log^2 m^2
\cr
&
+ \left[ \log^3\left(s_{1} s_{2}\over s_{4}\right)
- \frac{\pi^2}{3} \log\left(s_{1}\over s_{2} s_{4}\right)
+ 4 \log\left(s_{1}\over s_{4}\right)\,\text{Li}_2\left(1 - \frac{s_{1}}{s_{4}}\right)
\right.
\cr
&\left.
+ 4 \log\left(s_{1} s_{2}\over s_{4}\right)\,\text{Li}_2\left(1
- \frac{s_{2}}{s_{4}}\right)
- 4\,\text{Li}_3\left(1
- \frac{s_{2}}{s_{4}}\right)
+ 8 H_{011}\left(1 - \frac{s_{1}}{s_{4}}\right)
- 4 \zeta_3
\right] \log m^2
\cr
& + {\cal O}(\log^0 m^2),
\end{split}
\end{equation}
where we use
the harmonic polylogarithm function~\cite{Remiddi:1999ew}
\begin{equation}
H_{011}(1-x) = \frac{1}{2} \log(1-x) \log^2 x + \log x\,
\text{Li}_2(x) - \text{Li}_3(x) + \zeta_3.
\end{equation}
Finally, the pentagon-box integral $I^{(2)b}_5$ has the small $m^2$ expansion
\begin{equation}
\begin{split}
s_{2} s_{3} s_5 I^{(2)b} =&
- \frac{3}{4} \log^4 m^2
+ \left[
\log(s_{1} s_2 s_3 s_4)
- \frac{1}{3} \log(s_2 s_3 s_5)
\right] \log^3 m^2
\cr
& +
\left[
-\frac{3}{2} \log ^2s_{1}-2 \log s_2 \log s_{1}-\log s_3 \log s_{1}+2 \log s_4 \log s_{1}\right.\cr
& \left.+\log s_5 \log
    s_{1}+\frac{3}{2} \log ^2s_2+\log ^2s_3-\log ^2s_4+\frac{3}{2} \log ^2s_5\right.
\cr
&
\left.+\log s_2 \log s_3-2 \log s_2 \log
    s_4-2 \log s_3 \log s_4\right.\cr
&\left.-2 \log s_2 \log s_5-2 \log s_3 \log s_5+\log s_4 \log s_5
\right.
\cr
&\left.
-\,\text{Li}_2\left(1 - \frac{s_{1}}{s_3}\right)
+\,\text{Li}_2\left(1 - \frac{s_2}{s_4}\right)
-2\,\text{Li}_2\left(1 - \frac{s_2}{s_5}\right)
-2\,\text{Li}_2\left(1 - \frac{s_3}{s_5}\right)
+ \frac{5 \pi^2}{6}
\right]
\log^2 m^2
\cr
&+ \left[
-2 H_{011}\left(1 - \frac{s_3}{s_{1}}\right)
+8 H_{011}\left(1 - \frac{s_4}{s_{1}}\right)
-6 H_{011}\left(1 - \frac{s_4}{s_2}\right)
-4 H_{011}\left(1 - \frac{s_3}{s_5}\right)
\right.
\cr
&\left.
-8 H_{011}\left(1 - \frac{s_5}{s_2}\right)
- 4\,\text{Li}_3\left(1 - \frac{s_3}{s_1}\right)
+ 4\,\text{Li}_3\left(1 - \frac{s_4}{s_2}\right)
- 4\,\text{Li}_3\left(1 - \frac{s_3}{s_5}\right)
\right.
\cr
&\left.
+ 4\,\text{Li}_3\left(1 - \frac{s_5}{s_2}\right)
+ 2 \log\left(s_1 s_2 s_3\over s_4 s_5\right)
\,\text{Li}_2\left(1 - \frac{s_1}{s_3}\right)
+ 4 \log\left(s_1\over s_4\right)
\,\text{Li}_2\left(1 - \frac{s_1}{s_4}\right)
\right.
\cr
&\left.
+ 2 \log\left(s_1 s_4 s_5\over s_2^3 s_3\right)
\,\text{Li}_2\left(1 - \frac{s_2}{s_4}\right)
- 4 \log\left(s_2^2\over s_4 s_5^2\right)
\,\text{Li}_2\left(1 - \frac{s_2}{s_5}\right)
\right.\cr
&\left.
+ 4 \log\left(s_1 s_5\over s_3\right)
\,\text{Li}_2\left(1 - \frac{s_3}{s_5}\right)
+ \frac{8}{3} \log ^3s_1+2 \log s_2 \log ^2s_1-6 \log s_4 \log ^2s_1\right.\cr
&\left.
-\log s_5 \log ^2s_1+\log ^2s_2
    \log s_1+5 \log ^2s_4 \log s_1-\log ^2s_5 \log s_1\right.
\cr
&\left.
-2 \log s_2 \log s_4 \log s_1+2 \log s_3 \log
    s_5 \log s_1-\frac{1}{3} \pi ^2 \log s_1-\frac{17 }{3}\log ^3s_2\right.\cr
&\left.
-\log ^3s_3+3 \log ^3s_5-\log s_2 \log ^2s_3-4 \log
    s_2 \log ^2s_4+\log s_3 \log ^2s_4\right.\cr
&\left.-8 \log s_2 \log ^2s_5-2 \log s_3 \log ^2s_5-\log s_4 \log
    ^2s_5-\frac{2}{3} \pi ^2 \log s_2\right.
\cr
&\left.-2 \log ^2s_2 \log s_3-\frac{2}{3} \pi ^2 \log s_3+6 \log ^2s_2 \log s_4+2 \log
    s_2 \log s_3 \log s_4\right.\cr
&\left.
-\frac{1}{3} \pi ^2 \log s_4+9 \log ^2s_2 \log s_5+2 \log ^2s_3 \log s_5+2 \log s_2
    \log s_3 \log s_5\right.\cr
&\left.+\frac{1}{3} \pi ^2 \log s_5
+ 12 \zeta_3
\right] \log m^2
+ {\cal O}(\log^0 m^2)
\end{split}
\end{equation}
When we sum over cyclic permutations to obtain the full one-
and two-loop amplitudes,
the polylogarithm functions cancel, 
resulting in the relatively simple expressions~(\ref{fivepointoneloop})
and (\ref{fivepointtwoloop}).


\end{document}